\documentclass[notitlepage,superscriptaddress,aps,prl,twocolumn,noeprint]{revtex4-1}
\usepackage{amsmath}
\usepackage{amssymb}
\usepackage[sort&compress]{natbib}
\setcitestyle{numbers,comma,square}
\usepackage{upgreek}
\usepackage{mathtools}
\usepackage{graphicx}
\usepackage{afterpage}
\usepackage{amssymb}
\usepackage{epstopdf}
\usepackage{tikz}
\usepackage{verbatim}
\usepackage{lmodern}
\usetikzlibrary{arrows,shapes,backgrounds} 
\usepackage{tikzpagenodes} 

\tikzset{
    every picture/.style={remember picture},
    na/.style={baseline=-.5ex},
    background grid/.style={draw, black!50,step=.5cm}
}

\usepackage[colorlinks=true,citecolor=blue,linkcolor=magenta]{hyperref}
\usepackage{tikz}
\usepackage{color}
\usepackage{bm}
\usepackage[caption=false,position=top,singlelinecheck=off,justification=raggedright]{subfig}

\DeclarePairedDelimiter\bra{\langle}{\rvert}  
\DeclarePairedDelimiter\ket{\lvert}{\rangle}       
\DeclarePairedDelimiterX\braket[2]{\langle}{\rangle}{#1 \delimsize\vert #2}      
\def\lesssim{\ \raise.3ex\hbox{$<$}\kern-0.8em\lower.7ex\hbox{$\sim$}\ }
\def\gesim{\ \raise.3ex\hbox{$>$}\kern-0.8em\lower.7ex\hbox{$\sim$}\ }

\begin{document}
\title{Quantum Criticality in Open Quantum Spin Chains with Nonreciprocity}

\author{Samuel E. Begg}
\email{samuel.begg@apctp.org} 
\affiliation{Asia Pacific Center for Theoretical Physics, Pohang 37673, Korea}

\author{Ryo Hanai}
\email{ryo.hanai@yukawa.kyoto-u.ac.jp} 
\email{ryo.hanai@apctp.org}
\affiliation{Asia Pacific Center for Theoretical Physics, Pohang 37673, Korea}
\affiliation{Center for Gravitational Physics and Quantum Information, Yukawa Institute for Theoretical Physics, Kyoto University, Kyoto 606-8502, Japan}
\date{\today}

\begin{abstract}
We investigate
the impact of nonreciprocity on universality and critical phenomena in open quantum interacting many-body systems. Nonreciprocal open quantum systems often have an exotic spectral sensitivity to boundary conditions, known as the Liouvillian skin effect (LSE). 
By considering an open quantum XXZ spin chain that exhibits LSE, we demonstrate the existence of a universal scaling regime that is \textit{not} affected by the presence of the LSE. 
We resolve the critical exponents, which differ from those of free fermions, via tensor network methods and demonstrate that observables exhibit a universal scaling collapse, irrespective of the reciprocity. 
We find that the LSE only becomes relevant when a healing length scale $\xi_{\rm heal}$ at the system's edge (which is different to the localization length of the eigenstate of the Liouvillian)  exceeds the system size, allowing edge properties to dominate the physics. We expect this result to be a generic feature of nonreciprocal models in the vicinity of a critical point. The driven-dissipative quantum criticality we observe has no classical analogue and stems from the existence of multiple dark states.
\end{abstract}

\maketitle

\textit{Introduction ---}  
Universality in non-equilibrium systems can be seen in numerous phenomena ranging from directed percolation \cite{Hinrichsen2000}, flocking \cite{Vicsek1995,Toner1995,Benoit2019},  
Kardar-Parisi-Zhang physics \cite{Kardar1986} observed in various platforms \cite{Spitzer1970,Kriecherbauer2010,Takeuchi2010,Takeuchi2018,Ljubotina2017,Ljubotina2019,Scheie2021,Ilievski2021,Ye2022,Wei2022,Fontaine2022}, and nonreciprocal phase transitions \cite{Fruchart2021,Hanai2019,Hanai2020,Zelle2023,Chiacchio2023,You2020,Saha2020}. Recent advances in open quantum systems offer an exciting avenue for extending this concept. This is exemplified by criticality in non-Hermitian systems \cite{Ashida2017, Nakagawa2018}, open quantum systems \cite{Mitra2006,Prosen2008,Diehl2010,Dalla2010,Nagy2011,Oztop2012,Dalla2013,Sieberer2013,Cai2013,Marcuzzi2014,Tauber2014,He2015,Buchhold2015,Maghrebi2016,Marino2016prl,Marino2016prb,Sieberer2016,Foss2017,Nigro2019,Young2020,Hanai2020,Fujimoto2022,Zelle2023} and measurement-induced phase transitions \cite{Skinner2019,Fisher2023}. Engineered nonreciprocal couplings provide a promising direction for furthering these investigations. A number of platforms \cite{Fang2017,Gou2020,Xiao2020,Xiao2021,Lin2022,Liang2022,Lin2022magnet}, including an optomechanical circuit \cite{Fang2017} and cold-atoms \cite{Gou2020,Liang2022}, have demonstrated that asymmetric (nonreciprocal) transport can be engineered \cite{Metelmann2015,Song2019,Minganti2019,Liu2020,Longhi2020,Wanjura2020,Okuma2021,Haga2021,Wanjura2021,Mcdonald2022,Yang2022,Lee2022,Hu2023}.  Surprisingly, the spectrum of a system of particles hopping asymmetrically on a lattice  exhibits extreme sensitivity to changes in the boundary conditions, known as the non-Hermitian skin effect \cite{Lee2016,Yao2018,Kunst2018,Xiong2018,Mcdonald2018,Herviou2019,Borgnia2020,Okuma2020,Zhang2022,Gong2018,Kawabata2019,Zhou2019,Kawabata2019b,Ashida2020} or  Liouvillian skin effect (LSE) in the context of open quantum systems \cite{Song2019,Haga2021,Yang2022,Lee2022,Hu2023}. As spectral properties are usually a key element in determining the behavior of observables, one might expect that the presence of the LSE drastically alters the physics and hence the universal features (as was indeed shown in several works \cite{Haga2021,Kawabata2023,Hu2023}).

In this Letter, we introduce a nonreciprocal open quantum spin system that exhibits universal properties that are \textit{unaffected} by the LSE. For equilibrium critical phenomena, generic observables follow a universal power law as a function of the distance to the critical point $|T-T_c|$ (e.g., $C_{\rm V}\sim |T-T_c|^{-\alpha}$, where $C_{\rm V}$ is the specific heat and $\alpha$ is a critical exponent). This Letter considers an analogous situation in a nonreciprocal open quantum spin system exhibiting a 
quantum critical point.
As the parameter $\Gamma$ controlling the distance to the critical point is reduced, we observe universal behavior (e.g., $M\sim \Gamma^\alpha+\mathrm {const.}$, where $M$ is the magnetization) that is \textit{independent} of the strength of the nonreciprocity, contrary to the expectation given by a strong spectral sensitivity to boundary conditions. 
Quantum criticality
is demonstrated by the scaling collapse of observables, which exhibit the same critical exponents across various microscopic parameters. The resolved critical exponents differ from free systems, which we attribute to many-body interaction effects. We find that the LSE only impacts the bulk physics in the regime where the  healing length $\xi_{\rm heal}$ at the edge of the system (that diverges at the critical point $\Gamma\rightarrow 0$) exceeds the size of the system. 
This length scale $\xi_{\rm heal}$ is different from the localization scale of the eigenmodes of the Liouvillian \cite{Mcdonald2022}. We expect LSE-independent universality to be generic for nonreciprocal systems near a critical point.

\begin{figure}[t]
    \centering
 \subfloat{ \includegraphics[width=0.7\columnwidth]{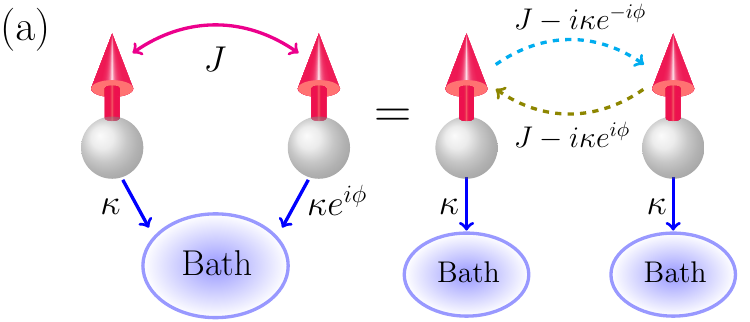}\label{fig:diagram}} 
   
      \vspace{-0.41cm}
   \subfloat{\includegraphics[width = 0.95\columnwidth]{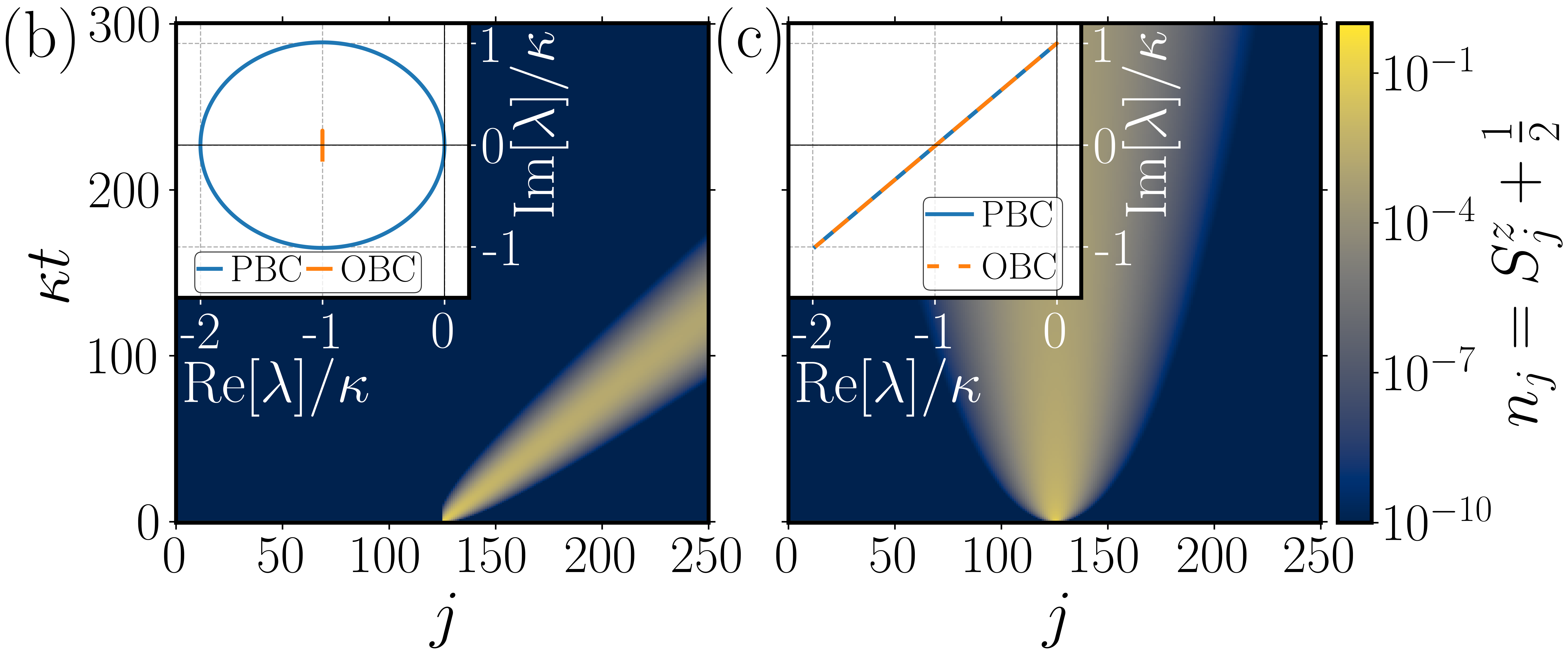} \label{fig:propagation_example}}
    \caption{
    (a) Left: two quantum spins coupled to a bath with coupling strength $\kappa$. The bath acts on the spins as $\hat{L}^l_{j}  = \sqrt{\kappa}( \hat{S}^-_j +  e^{i\phi} \hat{S}^-_{j+1} )$. 
    Right: the phase $e^{i\phi}$ causes interference that results in an effective system of nonreciprocally interacting spins and additional local on-site baths. (b) Relaxation of magnetization from the initial state with a single up-spin in the center of the chain for a nonreciprocal ($\phi = -\pi/2$) XX spin chain ($\Delta = 0$) with open boundary conditions. (c) The same for the reciprocal case ($\phi = 0$). 
 Insets: the spectrum $\lambda$ of the Liouvillian ${\mathcal L}$ in the single-magnon sector for periodic and open boundary conditions. We set $J/\kappa=1$ \cite{Fig1footnote}.
   }
    \label{fig:asymprop}
    
\end{figure}

\textit{Critical dynamics with nonreciprocity ---} 
To study the effect of nonreciprocity on universality, we consider a quantum spin system whose interactions are reservoir-engineered to be nonreciprocal. The evolution of the system's density matrix $\hat{\rho}$ in the presence of Markovian dissipation 
obeys 
the Lindblad master equation \cite{Lindblad1976}
\begin{align}
\frac{d\hat{\rho}(t)}{dt} & = \mathcal{L} [\hat{\rho}] =  -i [\hat{H}, \hat{\rho}(t)] +  \sum_j   \hat{\mathcal{D}}_j[\hat{\rho}(t)], \label{eq:lindblad}
\end{align}
with  dissipators 
$\hat{\mathcal{D}}_j[\cdotp] =  \hat{L}_{j}[\cdotp] \hat{L}^{\dagger}_{j}   - \frac{1}{2} \{\hat{L}^{\dagger}_{j} \hat{L}_{j}, [\cdotp] \}$ at site $j$.
 We solve Eq. (\ref{eq:lindblad}) using time-evolving block decimation (TEBD) \cite{Vidal2004,Schollwock2011,Paeckel2019}; see Supplemental Material (SM) for details \cite{LSE_SuppMat} and Refs. \cite{Zwolak2004,Verstraete2004open,Prosen2009,De2013,Daley2014,Cui2015,Mascarenhas2015,De2016,Gangat2017,Kshetrimayum2017,Gillman2019,Weimer2021} for examples of tensor networks applied to open quantum systems. We focus on the paradigmatic quantum XXZ spin Hamiltonian
\begin{align}
\hat{H} & = J \sum_j  \Big( \frac{1}{2} ( \hat{S}^-_j \hat{S}^+_{j+1} +   \hat{S}^+_j \hat{S}^-_{j+1} ) +  \Delta \hat{S}^z_j \hat{S}^z_{j+1} \Big) , \label{eq:XXZ}
\end{align}
where $J$ and $\Delta$ are the exchange interaction and anisotropy, respectively. The spin-1/2 operators obey 
$[\hat{S}_i^a,\hat{S}_j^b] = i \delta_{ij} \epsilon_j^{abc} \hat{S}_j^c$ and we set $\hbar = 1$. To study the nonreciprocal interaction effects, we use 
the dissipator
\begin{align}
    \hat{L}^l_{j} & = \sqrt{\kappa}( \hat{S}^-_j + e^{i\phi} \hat{S}^-_{j+1} ). \label{eq:dissipator} 
\end{align} 
In the SM \cite{LSE_SuppMat}, we provide a concrete proposal for implementing this correlated dissipation in a trapped ions platform using dissipative Aharanov-Bohm rings \cite{Metelmann2015, Clerk2022}, utilizing recent experimental advances \cite{Manovitz2020,Manovitz2022,Shapira2023}. 
It becomes clear that the dissipator \eqref{eq:dissipator} gives rise to a nonreciprocal interaction (Fig. ~\ref{fig:asymprop}(a)) by considering the conditional Hamiltonian,
$\hat{H}_{\text{cd}} = \hat H - \frac{i}{2}\sum_j \hat{L}^{l\dagger}_{j} \hat{L}^l_{j}$, governing the evolution in the absence of quantum jumps:
\begin{align}
    \hat{H}_{\text{cd}} \! = \!
  \! \sum_j \! \frac{J_+}{2}  \hat{S}^-_j \hat{S}^+_{j+1} \!+\! \frac{J_-}{2}\hat{S}^+_j \hat{S}^-_{j+1} \! + \!J \Delta \hat{S}^z_j \hat{S}^z_{j+1} \!\!- \! i  \kappa\hat{S}^+_j \hat{S}^-_j \! , \nonumber 
\end{align}
where $J_{\pm} = J - i e^{\mp i\phi}  \kappa$.  The phase factor $ e^{i\phi}$ in Eq. \eqref{eq:dissipator} therefore controls the nonreciprocity of interactions between nearest neighbor sites. The conditional Hamiltonian is similar to the non-Hermitian XXZ model considered in Ref. \cite{Albertini1996}. We stress, however, that we will investigate the unconditional dynamics including the effects of quantum jumps.

Figure \ref{fig:asymprop}(b) (Fig. \ref{fig:asymprop}(c)) demonstrates the anticipated nonreciprocal (reciprocal) transport of a spin excitation for $\phi=-\pi/2$ ($\phi=0$). Here, the spatial magnetization profile $S^z_j = \langle \hat{S}^z_j \rangle$ is plotted (with an offset for the ease of visibility), computed with open boundary conditions (OBC). A spectral sensitivity to boundary conditions (i.e., LSE \cite{Song2019,Haga2021}) in the nonreciprocal case is also observed (see insets of Figs.~\ref{fig:asymprop}(b),(c)), as expected.

 Interestingly, the relaxation of the system is far slower than the scales set by $\mathcal{O}(J^{-1})$ and $\mathcal{O}(\kappa^{-1})$ and is, in fact, algebraic (see Fig.~\ref{fig:chiral}), indicating that the system is critical. The slow relaxation occurs due to the presence of a dark state other than the all down state $\ket{\Downarrow} = \prod_j \ket{\downarrow}_j$.  To see this, let us temporarily assume a periodic boundary condition (PBC) and Fourier transform the dissipation terms in the Lindblad equation (\ref{eq:lindblad}), giving 
\begin{align}
 \sum_j   \hat{\mathcal{D}}_j[\hat{\rho}]  =    \sum_k  \kappa(k) \Big(\hat{S}^-_{k}\hat{\rho}   \hat{S}^{+}_{k}   - \frac{1}{2} \{\hat{S}^{+}_{k} \hat{S}^-_{k}, \hat{\rho}  \} \Big) , \label{eq:momemtlind}
\end{align}
where $\kappa(k) = 2 \kappa \big(1 + \cos(k + \phi )\big) $. Since the dissipator Eq.~\eqref{eq:dissipator} involves only spin flips from up to down, the system trivially possesses a dark state with all spins down $\ket{\Downarrow}$, i.e., $\mathcal {L}[\ket{\Downarrow}\bra{\Downarrow}]=0$. 
Notice, however, that the dissipation vanishes at $k = k^*= \pi -\phi$. This implies that a state $\ket{D_k}\equiv \hat{S}_k^+ \ket{\Downarrow}$  does not experience \textit{any} dissipation at $k=k_*$, where the operator $\hat{S}_k^+ =\frac{1}{\sqrt{L}}\sum_{j=1}^L e^{ikj} \hat{S}_j^+$ creates a spin-wave mode with momentum $k$. It can readily be shown that this is simultaneously an eigenstate of the Hamiltonian Eq.~\eqref{eq:XXZ}, which is a consequence of U(1) symmetry, making it a dark state $\mathcal {L}[\ket{D_{k_*}}\bra{D_{k_*}}]=0$~\cite{LSE_SuppMat}.
For $k$ very close to but not \textit{exactly} at $k=k_*$, $\ket{D_{k}}$ experiences a vanishingly small (but finite) dissipation rate \cite{DarkStatefootnote}. This implies that the characteristic time scale of the dissipation is divergent in the thermodynamic limit, meaning that the dynamics are critical, in agreement with Fig.~\ref{fig:asymprop}(b),(c).

The numerical results in Fig.~\ref{fig:asymprop}(b),(c) are obtained with OBC, while Eq.~\eqref{eq:momemtlind} is obtained under PBC. 
The two results are consistent with each other, despite the presence (absence) of the gap in the Liouvillian spectrum for OBC (PBC) [see Fig.~\ref{fig:asymprop}(b),(c) insets], because the local spin excitation will not know about the boundary conditions until they propagate or diffuse to hit the boundary \cite{Mcdonald2022, Lee2022}. This provides a key intuition: the \textit{spectral} sensitivity to boundary conditions does not necessarily imply the sensitivity for \textit{observables}.

In addition to the above-introduced engineered loss (Eq.~\eqref{eq:dissipator}), we further add a uniform gain to the system, $\hat L_j^g = \sqrt{\Gamma}\hat S_j^+$ \cite{FiniteSizefootnote}. This term invalidates the discussion above, introducing an additional time scale $\mathcal{O}(\Gamma^{-1})$ to the system.  Therefore, $\Gamma$ acts as a parameter that controls the distance from the critical point. Remarkably, despite the spectral sensitivity in the nonreciprocal case, which persists even for finite $\Gamma$ (see SM~\cite{LSE_SuppMat}), we will show that nonreciprocal and reciprocal systems display \textit{identical} universal properties in asymptotic regimes.

\begin{figure}[t]

     \subfloat{\includegraphics[width=\columnwidth]{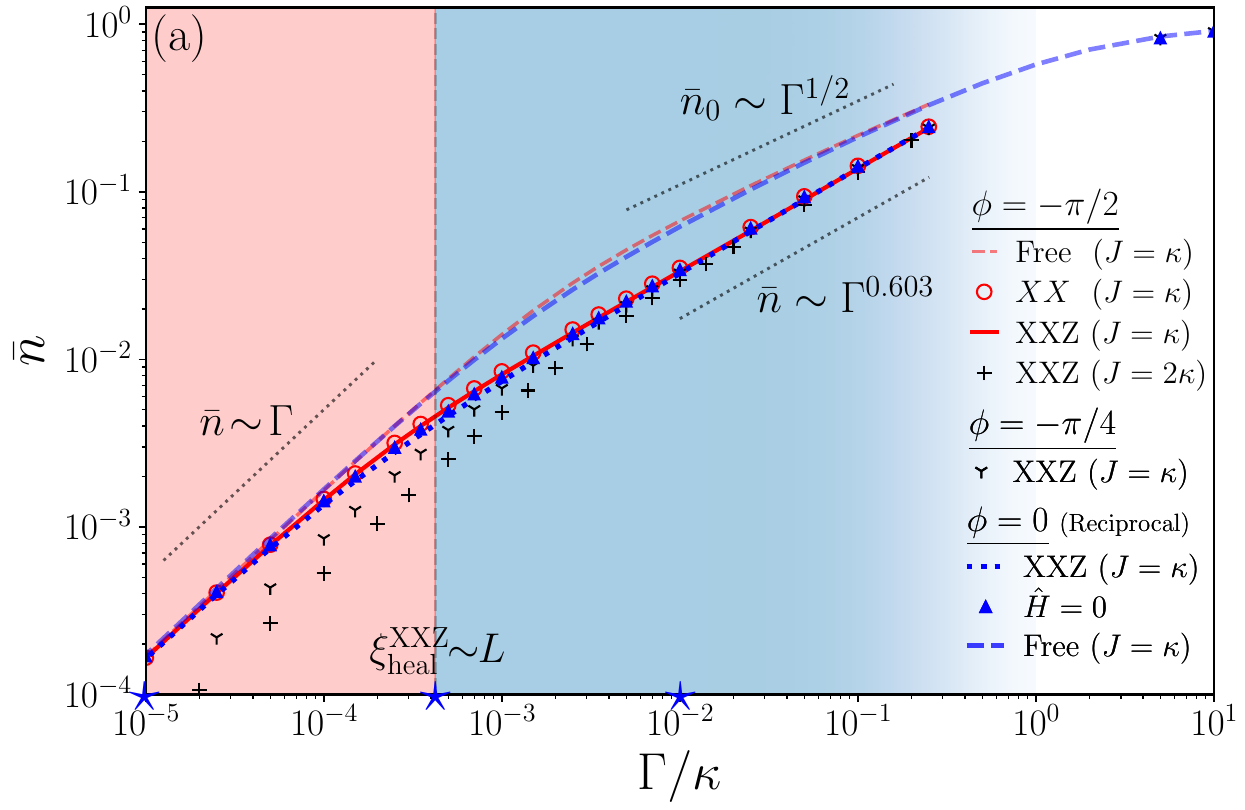} \label{fig:steadystate_a}}
    
     \vspace{-0.7cm}
   \begin{tikzpicture}
    \draw[->,black,semithick,dotted] (-10.9,-1.4) -- (-11,-0.8);
     \draw[->,black,semithick,dotted] (-8.9,-1.4) -- (-9.0,-0.6);
     \draw[->,black,semithick,dotted] (-5.7,-1.4) -- (-6.92,-0.74);
     \draw[->,white,thick] (-4.5,-1.4) -- (-6.7,-1.3);
   \end{tikzpicture}
   
       \vspace{-0.6cm}
    \subfloat{\includegraphics[width = 2.9cm]{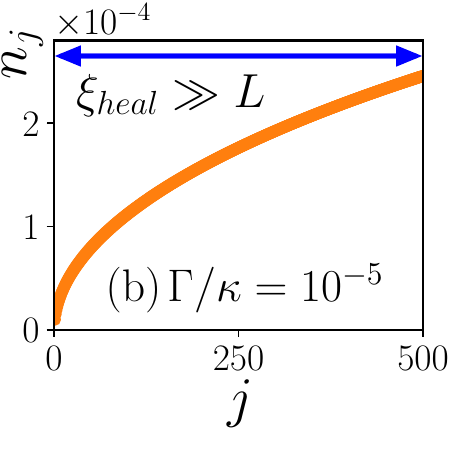}
   \hspace{-0.2cm}  \includegraphics[width = 2.9cm]{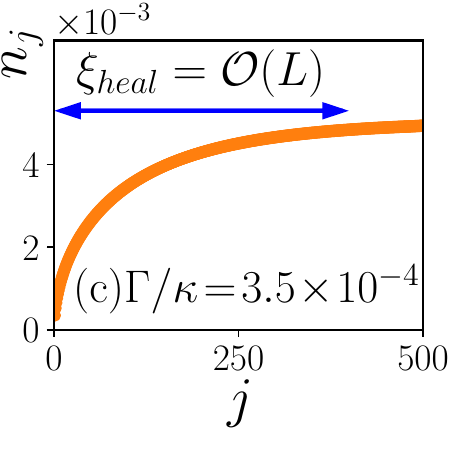}   \hspace{-0.2cm} 
    \includegraphics[width = 2.9cm]{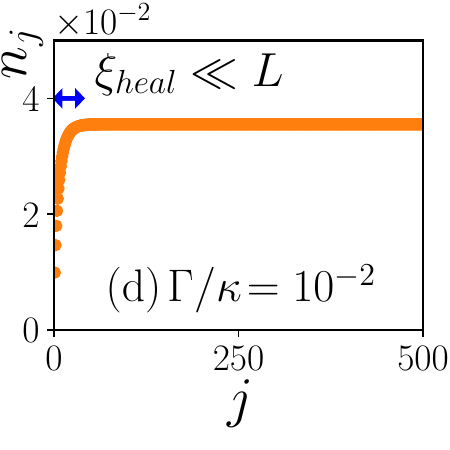}}
    \caption{
 (a)  Excitation density $\bar{n}$ vs $\Gamma$  in the steady state for different parameters.
Results are displayed for the following nonreciprocal ($\phi = -\pi/2$) systems:  XXZ ($\Delta= 2$), XX ($\Delta = 0$)  and free fermions for $L =500$, as well as XXZ with $J/\kappa = 2$ ($\Delta= 1$) for $L = 100$. For $\phi = -\pi/4$ we show XXZ ($\Delta =2$) with $L = 100 $.  Reciprocal ($\phi =  0$)  XXZ ($\Delta= 2$), $\hat{H}=0$, and free fermion results are also shown, all with $L = 50$.  Various fits $\bar{n} \sim \Gamma^{\alpha}$ are displayed (discussion in the text).     (b)-(d) Steady-state excitation density $n_j$ for different $\Gamma/\kappa$, corresponding to the red region $\xi_{\rm heal}\ll L$   (a),     the transition regime $\xi_{\rm heal}\sim L$ (b), and the asymptotic region $\xi_{\rm heal}\gg L$     (c),  respectively. The data corresponds to the nonreciprocal case with $\Delta = 0$ and $L = 500$. }
    \label{fig:steadystate}
\end{figure}
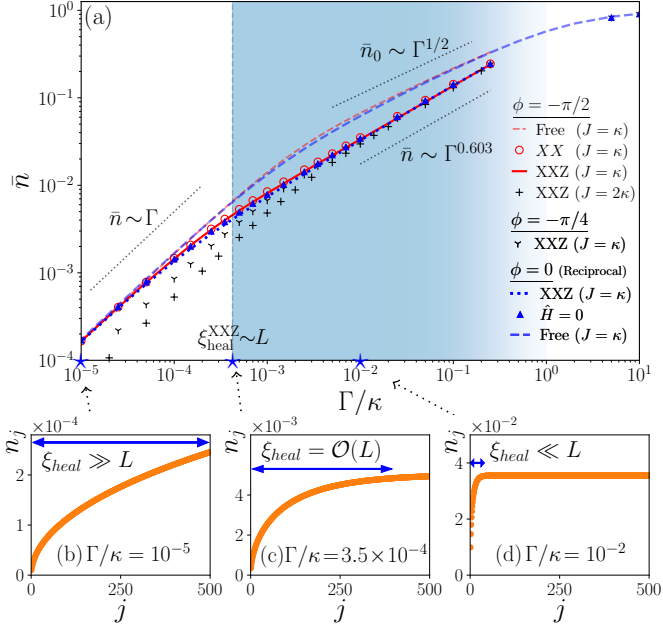

It is instructive to compare this model to a similar nonreciprocal free fermion model studied in Refs. \cite{Song2019,Wanjura2021,Mcdonald2022}
\begin{align}
\hat{H_{0}} \!= \!\!\sum_j \! \frac{J}{2} \big(\hat{c}^{\dagger}_j\hat{c} _{j+1}\! + \!\hat{c}^{\dagger}_{j+1}\hat{c} _{j}\big), \, 
\hat{L}_j^{l0} \!=\!\sqrt{\kappa}( \hat{c}_j + e^{i\phi}  \hat{c}_{j+1} ),\!\!\!
\label{eq:HatanoHop}
\end{align}
and $\!\hat L_j^{g0} = \sqrt{\Gamma}\hat{c}_j^\dagger$, where $\hat c_j$ is a fermionic annihilation operator satisfying $\{\hat{c}_i,\hat{c}_j^{\dagger} \} = \delta_{ij}$, and $ \{\hat{c}_i^{\dagger},\hat{c}_j^{\dagger} \} = \{c_i,c_j \} = 0$.
The conditional Hamiltonian $\hat H_{\rm cd}^0=\hat H_{0}-\frac{i}{2}\sum_j \hat L_j^{l0\dagger}\hat L_j^{l0}$ for this model ($\Gamma=0$ for simplicity) is given by the so-called Hatano-Nelson model \cite{Hatano1996,Hatano1997}, 
\begin{eqnarray}
    \hat H_{\rm cd}^0 = \sum_j \frac{J_+}{2}  \hat{c}_{j+1}^{\dagger}\hat{c}_j + \frac{J_-}{2} \hat{c}_j^+ \hat{c}_{j+1} - i \kappa  \hat{n}_j, \label{eq:fermcond}
\end{eqnarray}
where  $\hat{n}_j = \hat{c}^{\dagger}_j \hat{c}_j$ is the density operator. Eq. (\ref{eq:fermcond}) describes asymmetric hopping with an additional imaginary term.
A more direct comparison to our spin model can be made by performing the Jordan-Wigner transformation \cite{Coleman2015} for OBC, defined as
$ \hat{S}^+_j = e^{-i \pi \sum_i^{j-1}\hat{c}_i^{\dagger}\hat{c}_i}\hat{c}_j^{\dagger}, ~ \hat{S}^-_j =  e^{i \pi \sum_i^{j-1}\hat{c}_i^{\dagger}\hat{c}_i}\hat{c}_j, ~  \hat{S}^z_j =  \hat{n}_j -\frac{1}{2}.$
The jump operator (\ref{eq:dissipator}) and conditional Hamiltonian then take the form 
\begin{align}
 & \hat{L}^l_{j}  =  \sqrt{\kappa}( e^{i \pi \sum_i^{j-1}\hat{c}_i^{\dagger}\hat{c}_i}\hat{c}_j +   e^{i\phi} e^{i \pi \sum_i^{j}\hat{c}_i^{\dagger}\hat{c}_i}\hat{c}_{j+1} ),
\label{eq:jumpstring} \\
& \hat{H}_{\text{cd}}   = 
 \hat H_{\rm cd}^0 
 + J \Delta \Big(\hat{n}_j \hat{n}_{j+1}
 -  \hat{n}_j + \frac{1}{4} \Big),\label{eq:hcondspinJW}
\end{align}
where one sees that $\hat{H}_{\rm cd}$ is given by the Hatano-Nelson model~\eqref{eq:fermcond} extended to have nearest-neighbor interactions, suggesting that the free fermion model (Eq.~\eqref{eq:HatanoHop}) can be regarded as the non-interacting limit of our spin model and serves as a useful point of reference. Note that, while the string operators $e^{\pm i \pi \sum_i^{j-1}\hat{c}_i^{\dagger}\hat{c}_i}$ in the conditional Hamiltonian (\ref{eq:hcondspinJW}) have cancelled out, those  in the quantum jump term $\hat{L}^l_{j}\hat{\rho}(t) \hat{L}^{l\dagger}_{j}$ cannot be removed. This means that even the XX model case $\Delta=0$ does \textit{not} correspond to a free system.

\textit{Universality and scaling collapse ---}
Figure \ref{fig:steadystate}(a) shows the spatially averaged  excitation number $\bar{n} = \frac{1}{L}\sum_j n_j$ in the steady state as a function of $\Gamma$, where $n_j = \langle \hat{n}_j \rangle=\langle \hat{S}^z_j \rangle +\frac{1}{2}$. Here, data is shown for a variety of parameters, including different strength of nonreciprocity $\phi$, $\Delta$, $J$. Data for different system sizes is in the SM \cite{LSE_SuppMat}. For comparison, the free fermion case is also plotted. Consistent with the property that $\Gamma=0$ is a critical point, we observe the power-law scaling $\bar{n} \sim \Gamma^{\alpha}$. Remarkably, the exponent $\alpha=0.603(9)$  in the blue shaded region is identical in all cases, including both nonreciprocal ($\phi=-\pi/2$) and reciprocal ($\phi=0$) cases for XXZ ($\Delta>0$) and XX models ($\Delta=0$), different exchange interaction strengths $J$, and even a purely dissipative case ($\hat H=0$). The obtained exponent $\alpha=0.603(9)$ is different from the free fermion case $\alpha = 0.5$. The result clearly demonstrates that universal features have emerged, irrespective of the presence of the LSE. For the purely dissipative case ($\hat H=0$), $\phi$ can be removed from the Liouvillian (\ref{eq:lindblad}) via a local gauge transformation $\hat{S}^-_j \rightarrow e^{-i \phi j}\hat{S}^-_j$, further illustrating that the scaling is independent of reciprocity.

\begin{figure}[t]
    \centering
   \hspace{1cm} \includegraphics[width=0.55\columnwidth]{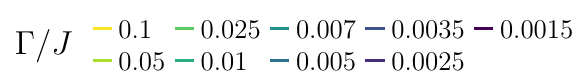}
   
      \vspace{-0.32cm}
      
     \subfloat{\includegraphics[width=\columnwidth]{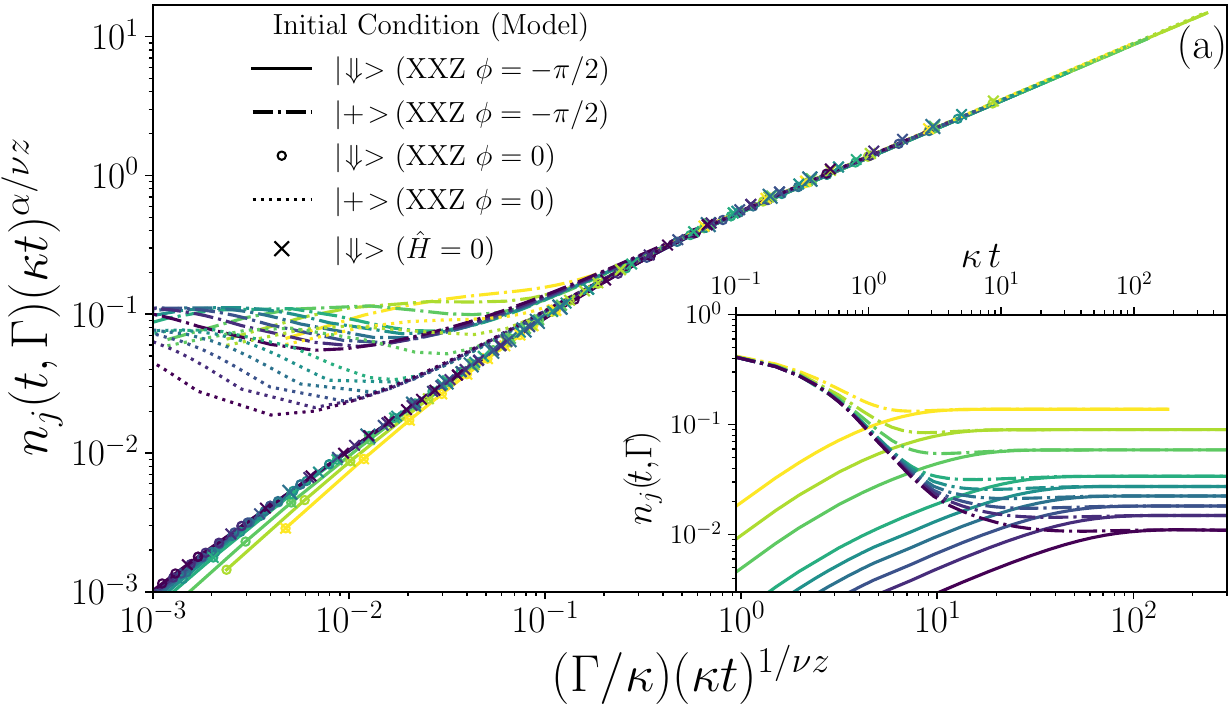}\label{fig:densitycollapse}}
     
       \centering
   
       \vspace{-0.1cm}
       
     \subfloat{\includegraphics[width=\columnwidth]{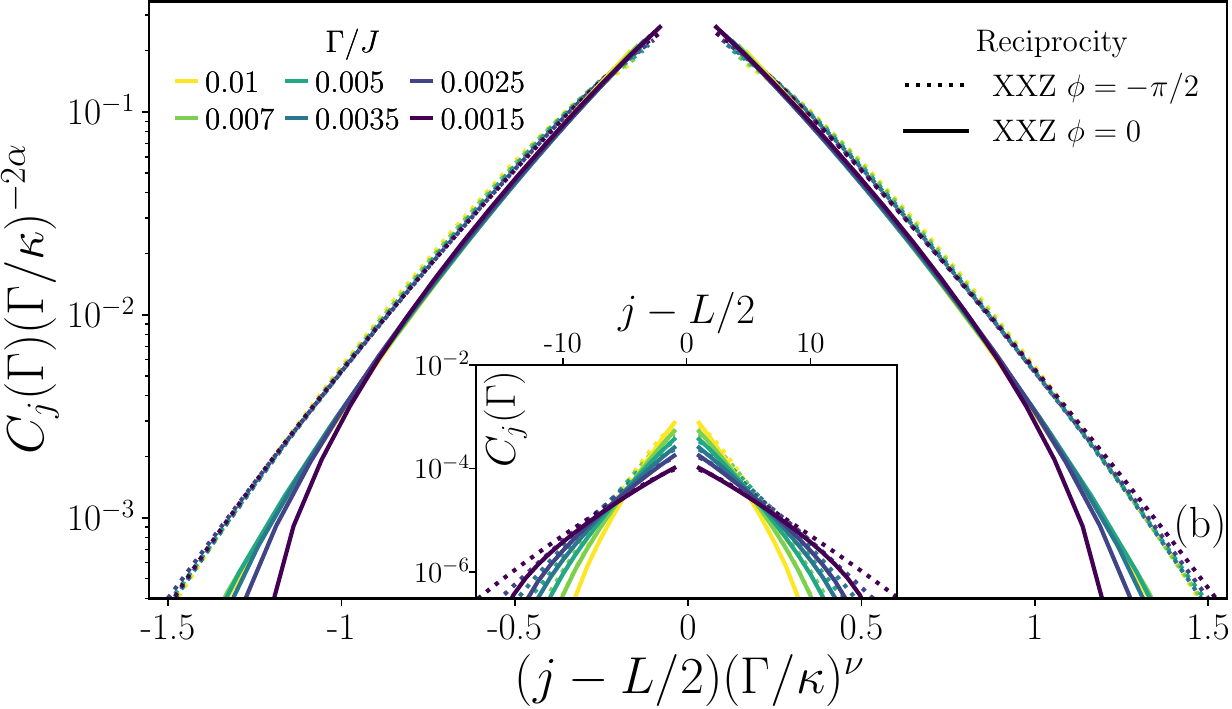} \label{fig:corr_collapse}} 
    \caption{ (a) $n_j(t,\Gamma)(\kappa t)^{\alpha/\nu z} $ vs $(\Gamma/\kappa)(\kappa t)^{1/\nu z}$
    with $\alpha=0.603$, $\nu = 0.386$ and $z = 1.96$ for the  $J/\kappa=1$, $\Delta = 2$, dissipative XXZ model over a range of $\Gamma$ values, 
    with  $L = 500$ and setting $j = 475$. Results are displayed for nonreciprocal ($\phi = -\pi/2$), reciprocal ($\phi = 0$), and $\hat{H} = 0$ cases, with initial conditions being the  
    fully-polarized state $\ket{\Downarrow}$ and the x-polarized state $\prod_j^L \ket{+}_j$ respectively. 
    The inset shows un-scaled data for the nonreciprocal cases. 
    (b)  Scaled connected correlation $C_j(\Gamma)(\Gamma/\kappa)^{-2\alpha}$ vs $(j-L/2)(\Gamma/\kappa)^{\nu}$ in the steady-state for a range of $\Gamma$ values.
   The inset shows the un-scaled data.
    }
    \label{fig:relaxation}
\end{figure}

Figure \ref{fig:relaxation} demonstrates a scaling collapse of the density and the spatial correlation function in this region:
\begin{eqnarray}
 n_j(t,\Gamma) 
 & = & t^{-\alpha/\nu z} f_{{n}_j}  (t\Gamma^{\nu z}  ) , \label{eq:densityscaling} \\
 C_j(\Gamma) &=&    \Gamma^{2\alpha} f_{{C}_j}(\Gamma^{\nu} (j-L/2)),  
 \label{eq:corrscale}
\end{eqnarray}
 where $C_j(\Gamma)$ is the magnitude of connected correlations between a site $j$ and the center of the chain $L/2$, $C_j(\Gamma)  =|\langle \hat{S}^z_{L/2}\hat{S}^z_{j}\rangle -\langle \hat{S}^z_{L/2}\rangle\langle  \hat{S}^z_{j}\rangle  |$. Here, $f_{{n}_j}(x),f_{{C}_j}(x)$ are scaling functions for the density and spatial correlation function,  respectively, while $\alpha$, $z$ and $\nu$ are critical exponents that characterize the universal features. Data is provided for reciprocal and nonreciprocal cases and for different parameters and initial states. The scaling collapse is achieved, by setting the critical exponents  $\{z,\nu,\alpha\}= \{1.96(13),0.386(16),0.603(9) \} $ \cite{Fittingfootnote}, unambiguously demonstrating the emergence of universality. 
For the free fermions we find
$\{z,\nu,\alpha\}= \{2,0.5,0.5 \}$ \cite{LSE_SuppMat}. 
The critical phenomenon we observe is similar to the quantum critical phenomena proposed in Refs. \cite{Marino2016prl,Marino2016prb} for a driven-dissipative bosonic system. However,  their system has a steady state that is interacting, while our steady state at the critical point is a vacuum, resulting in a different universality class characterized by critical exponents $\{z,\nu,\alpha\}= \{2.025,0.405,0.5\}$.

In the  regime of sufficiently small $\Gamma\lesssim J/L$ (the region shaded in red in Fig. \ref{fig:steadystate}(a)), we observe that the scaling properties change to $\bar{n} \sim \Gamma^{\alpha'}=\Gamma$, i.e. $\alpha'=1$. (Note however that this regime shrinks to measure zero as the system size is increased.) This can be understood from the steady state density profile $n_j$ in Fig. \ref{fig:steadystate}(b)-(d), which shows results for different $\Gamma$ values. As seen, the density profile exhibits a dip at the left boundary, with its healing length $\xi_{\rm heal}$ (characterizing the length of the dip) decreasing as a function of $\Gamma$. The dip arises because sites near the left boundary do not experience any flux of incoming  excitations from the boundary, whereas sites in the bulk are `topped up' from their left. As these spin waves exhibit an increasingly long lifetime as $\Gamma$ decreases, the healing length $\xi_{\rm heal}$ becomes increasingly long and diverges at $\Gamma\rightarrow 0$. Note that $\xi_{\rm heal}\propto\Gamma^{-1}$ is very different from the localization length $\xi_{\rm loc}$ of the eigenmodes of the Liouvillian $\mathcal {L}$, which is solely determined by the asymmetry of the hopping $\xi_{\rm loc}\sim 1/\log(|J_+|/|J_-|)$ \cite{Mcdonald2022}. 

In the asymptotic regime ($J/L\lesssim\Gamma\lesssim \kappa$)
(Fig.~\ref{fig:steadystate}(d)), the healing length $\xi_{\rm heal}$ is small compared to the system size $L$. 
Therefore, the density profile 
is almost uniform. 
As $\Gamma$ decreases to $\Gamma \lesssim J/L $ the healing length starts to exceed the system size (Fig.~\ref{fig:steadystate}(b),(c)). 
This implies that, while in the asymptotic region $J/L\lesssim\Gamma\lesssim \kappa$, the physics is determined by the \textit{bulk} properties (that do not care about LSE \cite{Okuma2021b, Mcdonald2022}), the region with $\Gamma \lesssim J/L$ is dominated by the \textit{edge} properties, giving a natural explanation for the change of scaling properties at different regimes. 
The scaling $\bar{n} \sim \Gamma$ is consistent with the free fermion case with perfect nonreciprocity ($\phi = -\pi/2$, $J=\kappa$) for $\xi_{\rm heal}\gg L$ \cite{Mcdonald2022}. Interestingly, while in this limit the transition between the two regimes occurs at $\Gamma = \mathcal{O}(v_g /L)$ for free fermions, for the spin systems, the many-body interaction alters the scaling to $\Gamma = \mathcal{O} (v_g /L^{1.25}),$ where $v_g= J \sin (\pi - \phi)$ is the group velocity of the least damped mode $k_*$.

\begin{figure}
    \centering
     \subfloat{
    \includegraphics[width=0.538\columnwidth]{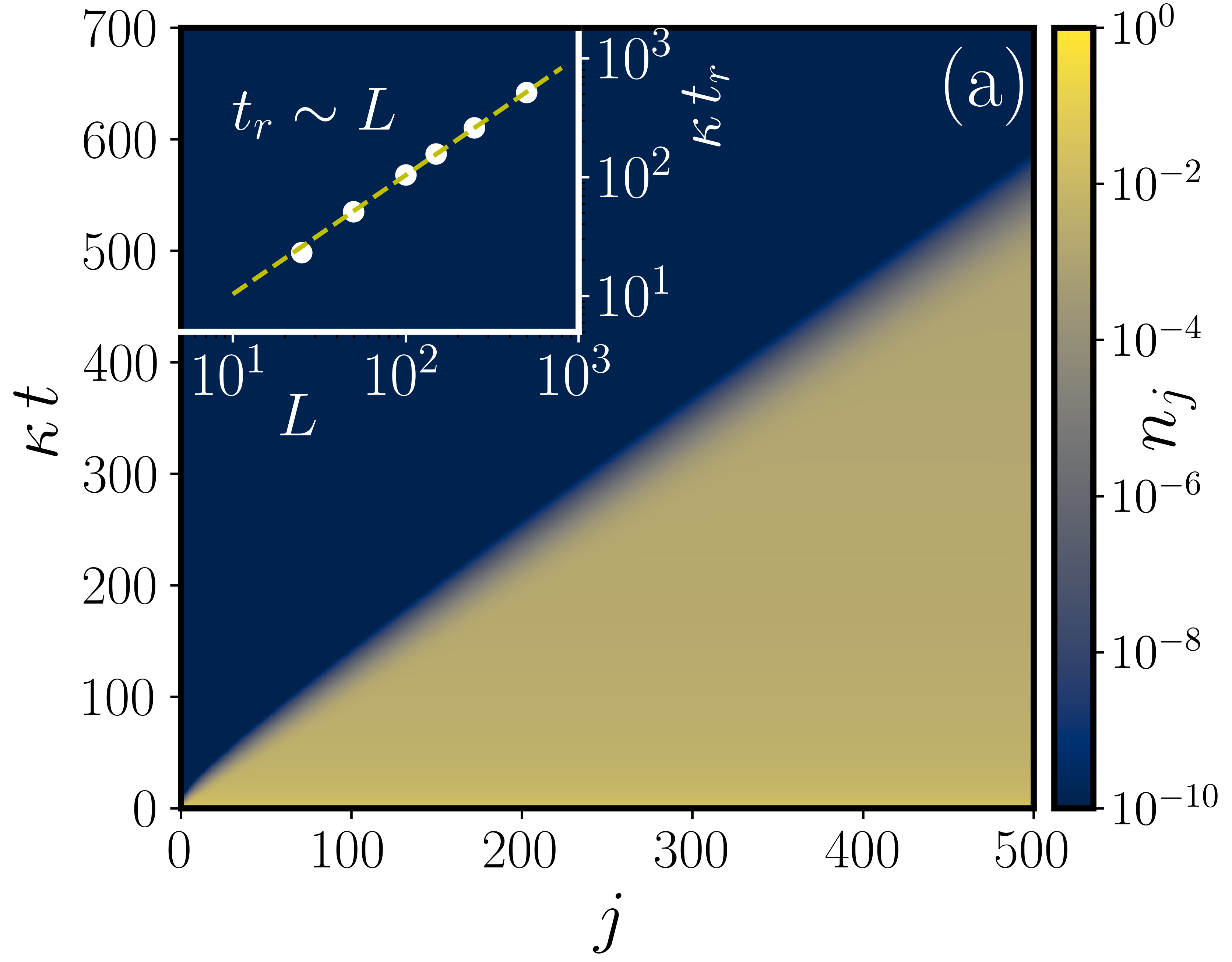}\label{fig:lse_spacetime}}
       \subfloat{  \includegraphics[width =0.462\columnwidth]{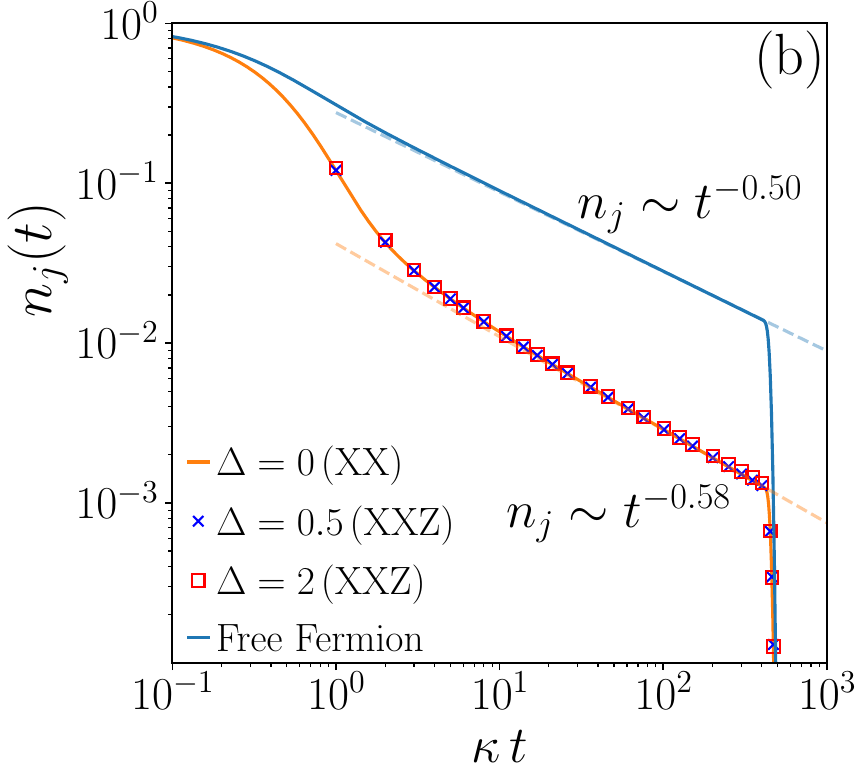} \label{fig:lse_powerlaw}}
        \caption{   (a) Space-time plot of the particle density $n_j$ for the nonreciprocal ($\phi = -\pi/2$) XXZ  spin chain with $L = 500$ sites,  starting from all spins up and with $\Gamma = 0$, $J/\kappa = 1$,  $\Delta= 2$.    Inset: the relaxation time to the steady-state, $t_r$, vs system size $L$, with linear fit $t_r \sim L$ (dotted).
    (b) Density decay $n_j$ vs time with $j = 450 $, for various system parameters.
    }
    \label{fig:chiral}
\end{figure}

The scaling properties in the  region $\Gamma \lesssim J/L$ 
are strongly affected by the LSE.  This is demonstrated in Fig \ref{fig:chiral}(a), which shows a spacetime plot of the excitation density for $\Gamma = 0$, starting from an initially fully-polarized state with all spins up $\ket{\Uparrow}.$ Here, the excitation density $n_j(t)$ 
exhibits a sudden transition from power law to exponential decay \cite{Song2019,Haga2021,Yang2022}. 
This occurs when site $j$ is no longer `topped up' by incoming excitations: all the long-lived excitations that were initially left of site $j$ have propagated to its right \cite{Lee2022}. 
For sites near the right edge, which are last to relax, this 
takes a time $t_r$ proportional to the system size, i.e. $ t_r \sim \xi^{z'} = L$  ($z' = 1$). In comparison, the transport is approximately
diffusive in the reciprocal case ($z' \approx 2$), which is clearly visible in Fig. \ref{fig:asymprop}(c). Therefore, in the region $\Gamma \lesssim J/L$ the scaling is altered by the LSE.  In the SM \cite{LSE_SuppMat}, we show for the free fermion system that under PBC this region only arises for $\Gamma \leq \mathcal{O} (1/L^2)$.

Finally, Fig. \ref{fig:chiral}(b) shows that many-body effects also alter the power-law exponent $\chi=0.58$ of $n_j(t) \sim t^{-\chi}$ at $\Gamma = 0$ 
from the free fermion result $\chi=0.5$. Curiously, $\chi$ is found to be initial-state-dependent ranging from $\chi=0.5$ to 0.58 \cite{LSE_SuppMat}. 
Clarifying the origin of this remains our future work.

\textit{Discussion---} 
In conclusion, we have demonstrated the existence of a
Liouvillian skin effect (LSE)-independent
universal regime.
We showed that the LSE can affect the bulk properties only 
when $\xi_{\rm heal}\gesim\mathcal{O}(L)$. 
The LSE-induced transition of scaling reported in Refs.~\cite{Haga2021, Kawabata2023}  
corresponds to the latter regime 
where the number conservation in their model implies the absence of characteristic length scales (similar to ``model B'' of Ref.~\cite{Hohenberg1977}).

\textit{Acknowledgements ---} 
We thank Aashish Clerk, Kazuya Fujimoto, Tomohiro Sasamoto, and Hironobu Yoshida for useful discussions and Hosho Katsura and Alexander McDonald for the critical reading of the manuscript. This work was supported by an appointment to the JRG program at the APCTP through the Science and Technology Promotion Fund and Lottery Fund of the Korean Government, as well as by Grant-in-Aid for Research Activity Start-up from JSPS in Japan (Grant No. 23K19034), and by the National Research Foundation (NRF) funded by the Ministry of Science of Korea (Grant No. RS-2023-00249900). SEB acknowledges the support of the Young Scientist Training Program at the Asia Pacific Center for Theoretical Physics.
 

\begin{thebibliography}{123}
\bibitem [{Fig()}]{Fig1footnote}%
  \BibitemOpen
  \href@noop {} {}\bibinfo {note} {For ease of visibility, the insets have
  $\kappa = 1.01 J$ \cite{Mcdonald2022} (rather than $J/\kappa = 1$ as in the
  main plots), since the spectrum for open boundary condition reduces to a
  single point $\lambda = -J$ when $J= \kappa$.}\BibitemShut {Stop}%
\bibitem [{Dar()}]{DarkStatefootnote}%
  \BibitemOpen
  \href@noop {} {}\bibinfo {note} {We note that a generic initial condition
  relaxes to an all down state $\ket{\Downarrow}$ rather than the spin-wave
  dark state $\ket{D_{k_*}}$, which can be understood as follows. Since there
  can be no transitions between the single magnon eigenstates, the only way to
  reach the dark state $\ket{D_{k_*}}$ is via quantum jumps from the two magnon
  sector. There are $L$ linearly independent two magnon states that satisfy
  this requirement (see SM \cite{LSE_SuppMat}). The remaining
  $\mathcal{O}(L^2)$ basis states will decay to the dark state with all spins
  down $\ket{\Downarrow}$, which is therefore the steady state for generic
  initial states at large $L$. Nevertheless, the dark state $\ket{D_{k_*}}$
  plays a dynamical role that determines the asymptotic relaxation towards the
  all down state $\ket{\Downarrow}$.}\BibitemShut {Stop}%
\bibitem [{Fin()}]{FiniteSizefootnote}%
  \BibitemOpen
  \href@noop {} {}\bibinfo {note} {To reduce the impact of finite size effects,
  additional jump operators $\hat L_{1/L} = \sqrt{\kappa}\hat S_{1/L}^-$ are
  applied on the boundary sites, which ensures that the model has the same
  magnitude of on-site spin sinks (and sources) on all sites.}\BibitemShut
  {Stop}%
\end{thebibliography}
\bibliographystyle{apsrev4-1}

\end{document}


\title{Supplemental Material for ``Quantum Criticality in Open Quantum Spin Chains with Nonreciprocity"}

\author{Samuel E. Begg}
\affiliation{Asia Pacific Center for Theoretical Physics, Pohang 37673, Korea}
\author{Ryo Hanai}
\affiliation{Asia Pacific Center for Theoretical Physics, Pohang 37673, Korea}
\affiliation{Center for Gravitational Physics and Quantum Information, Yukawa Institute for Theoretical Physics, Kyoto University, Kyoto 606-8502, Japan}

\date{\today}

\maketitle

\subsection{Experimental Proposal}
\begin{figure}[t]
\subfloat[]{\includegraphics[width = 3.5cm]{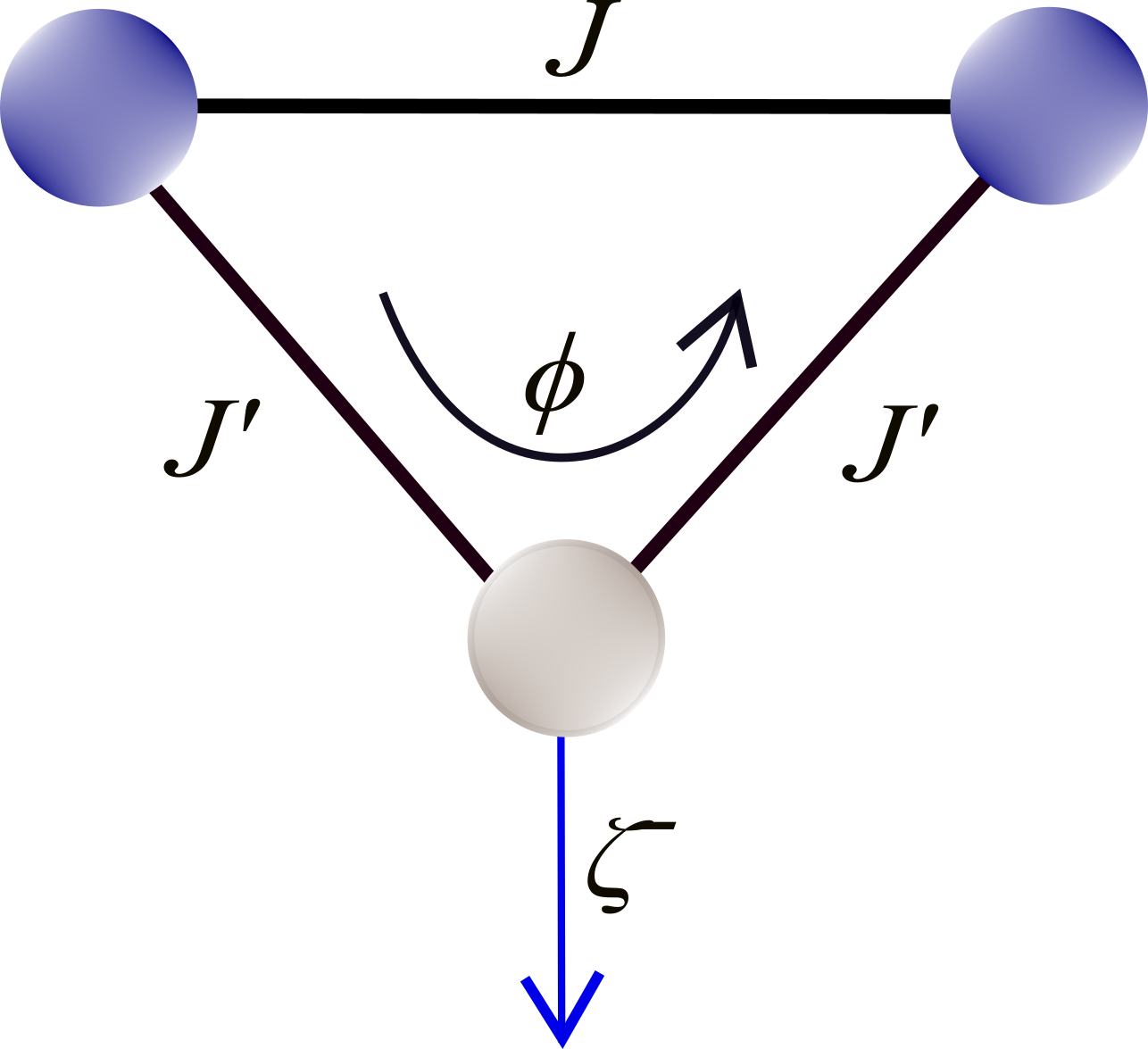}} \hspace{0.5cm}
\subfloat[]{\includegraphics[width = 3.5cm]{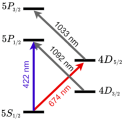}}

\subfloat[]{\includegraphics[width = 8.7cm]{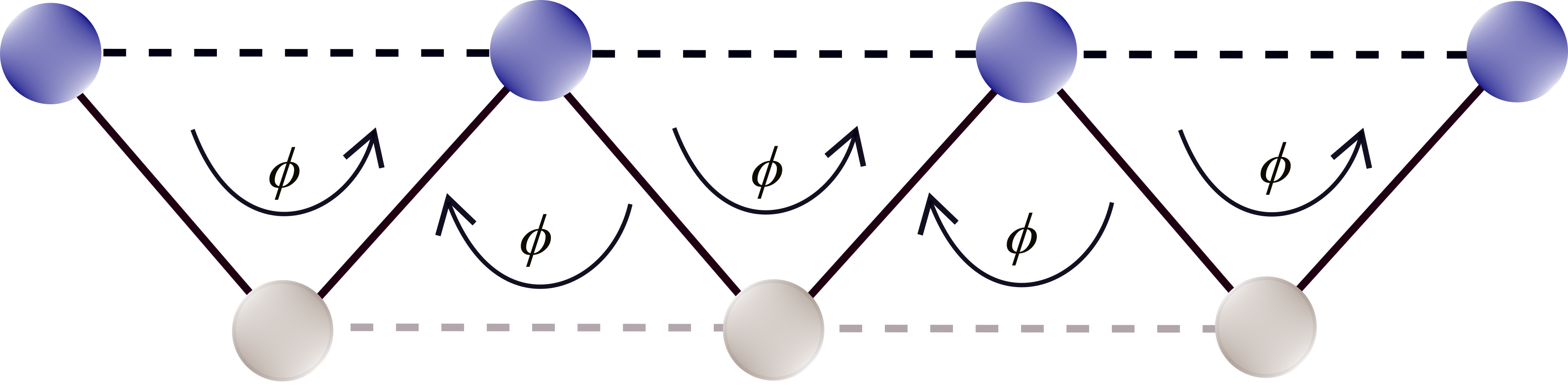}}

\subfloat[]{\includegraphics[width = 8.7cm]{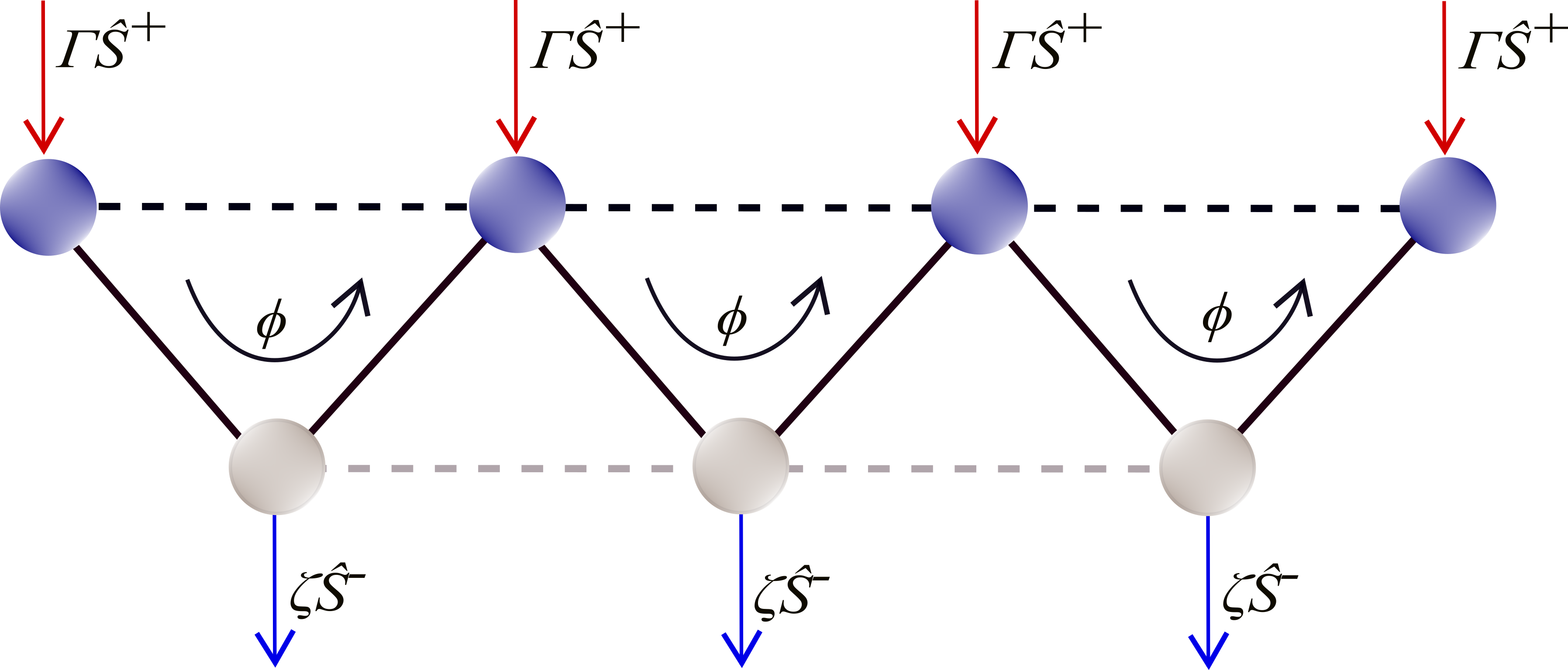}}

\caption{(a) Basic ingredients needed to engineer an effective nonreciprocally interacting system: hopping in a three-site system with rates $J$ and $J'$, an Aharanov-Bohm phase $\phi$ and on-site particle loss with rate $\zeta.$  (b) Energy level diagram of Strontium ions, together with transition wavelengths, adapted from Ref. \cite{Manovitz2022} licensed under a Creative Commons Attribution 4.0 International License (\href{https://creativecommons.org/licenses/by/4.0/legalcode}{CC BY 4.0}). A qubit can be encoded in long-lived $5S_{1/2}$ and $4D_{5/2}$ orbital states, while $5P$ states are short-lived and can be used to mediate optical pumping and loss. (c) Ladder geometry constructed from Aharanov-Bohm rings. Nearest-neighbour interactions in the 1D ion chain are indicated by the solid dark lines, while next-nearest-neighbour interactions are indicated by dashed lines. (d) Diagram of the proposed set-up for a trapped ions implementation of a nonreciprocally interacting spin chain, obtained by adding optical pumping and loss to (c). The nonreciprocally interacting sites are shaded dark, whereas auxiliary sites that mediate this interaction are shaded light.  Interactions between auxiliary sites (dashed and faded lines) do not contribute significantly once dissipation is introduced, see discussion in text. }
\label{fig:experimentaldiagram}
\end{figure}

Here we propose a trapped ion implementation of the spin model, utilizing the recent observation of Aharanov-Bohm rings using Strontium ions \cite{Shapira2023}. 
Before introducing this, we 
briefly review
how to implement nonreciprocity in the case of free particle systems of fermions or bosons \cite{Metelmann2015,Clerk2022}.
 Nonreciprocal interactions between two sites
 can
 be mediated by an auxiliary non-equilibrium degree of freedom. 
 Figure \ref{fig:experimentaldiagram}(a) illustrates the simplest case of a system 
 composed of a three-site ring exposed to a gauge flux $\phi$, with the aim of engineering nonreciprocity between two of the sites \cite{Clerk2022}. 
The Hamiltonian for this system
is given by
 \begin{align}
\hat{H}  = -e^{i\phi/3}(J c^{\dagger}_{1}c_{2} + J' c^{\dagger}_{2}c_{3} + J' c^{\dagger}_{3}c_{1}) + {\rm h.c.},
 \end{align}
Here, $J$ is the hopping rate between the sites of interest, sites 1 and 2, while $J'$ is the rate for hops involving the auxiliary site 3.  
In addition, we also add
laser-induced single-site loss described by a jump operator $\hat{L} = \sqrt{\zeta} \hat{c}_3$. 
Nonreciprocity arises via the interference of various paths within the ring.
To see this, let's consider the possible
paths a particle may take from site 2 to site 1. 
The particle can either hop directly from the site 2 to 1 or can also hop from 2 to 3, and then to 1 (up to second order in the hopping).
The associated 
 probability amplitude $G^R[j,j',\omega]$ to go from site $j'$ to $j$ is given by 
 \begin{align}
  G^R[1,2,\omega] & =   -J e^{i\phi/3} + (-J' e^{-i\phi/3})\frac{1}{\omega - i \zeta/2} (-J' e^{-i\phi/3}) \nonumber \\ & + \mathcal{O}(J^3),
 \end{align}
where we have assumed $J$ and $J'$ are real without loss of generality.
  The first term is due to the direct hop from site 2 to 1, while the second term is due to the indirect path via site 3, with a contribution from the associated dissipation on this site. 
One may tune this amplitude to zero, i.e., 
$G^R[1,2,\omega]=0,$ 
by setting
\begin{align}
 \zeta/2 = \sqrt{(J'^{2}/J)^2-\omega^2},~~~~ \tan \phi = \frac{\zeta}{2\omega} .\label{eq:interference_conditions}
 \end{align}
 The presence of the particle loss $\zeta>0$ ensures the reverse process $G^R[2,1,\omega]$ is non-zero. Rather, in the latter, the interference is constructive \cite{Clerk2022}, leading to enhanced hopping. Therefore, this mechanism successfully implements nonreciprocal propagation.  
 
When we set
 the loss on the auxiliary site 
to be
 large compared to the response frequency $\zeta \gg \omega$, which amounts to an adiabatic approximation \cite{Clerk2022}, Eq. (\ref{eq:interference_conditions}) reduces to $\zeta = 2J'^{2}/J$ and $\phi = \pi/2.$ 
 The resulting jump operator is given by $\hat{L} = \sqrt{\kappa} (\hat{c}_1 + e^{i\phi}\hat{c}_2),$ where $\kappa/2 = 2J'^2/\zeta = J$ \cite{Clerk2022}. 
A set of consistent scales is given by $\zeta \gg J' \gg \kappa = \mathcal{O}(J)$. 
 The nonreciprocity can then be easily understood as arising via interference between the Hamiltonian component $\hat{H}$ and dissipative terms in the Lindblad equation according to the discussion in the main text.  
By introducing multiple three site rings a chain of sites with nonreciprocal hopping can be designed.
These principles were utilized in recent cold atom implementations of nonreciprocity \cite{Gou2020, Liang2022}.

We wish to propose a spin analog of this free fermion system. We utilize
the methods proposed and experimentally implemented in Ref. \cite{Shapira2023} using trapped ions, which are built on recent advances \cite{Manovitz2020, Manovitz2022}. Ref. \cite{Shapira2023} successfully implemented a spin system described by the Hamiltonian,
\begin{align}
\hat{H}  = \sum_n \Omega_n e^{i(\phi_n - \delta_n t)} \sum_{j} \hat{\sigma}^{\dagger}_{j}\hat{\sigma}_{j+n} +   {\rm h.c.}, \label{eq:experimental_ham}
 \end{align}
where $\phi_n,$ $\delta_n$ and $\Omega_n$ are all experimentally tunable parameters. 
For our purpose, it is enough to focus
on the case in which $\delta_n = 0$. 
This Hamiltonian corresponds to a hopping model with 
 hopping amplitude
$\Omega_n$
 to a site that is $n$ sites apart. 

Let us briefly summarize how this was realized experimentally in Ref. \cite{Shapira2023}. A M{\o}lmers-S{\o}renson interaction \cite{Molmer1999}  was applied to a trapped-ion chain, with additional drive detuning to remove pair creation and annihilation processes $\sigma^+_i \sigma^+_j$ and $\sigma^-_i \sigma^-_j$. The remaining interactions are of the form $\sigma^+_i \sigma^-_j$. However, since the energies of all ions are degenerate, all-to-all hopping is generated. A magnetic field gradient is then applied to the chain, which detunes the qubit transition frequencies for each ion, severing the all-to-all coupling. The excited state of neighbouring ions then differs by a frequency $f$, and $nf$ for ions spaced at a distance of $n$ sites. The hopping is selectively reinstated by generating spectral tones with a frequency difference $nf$. The phase difference between the tones gives the phase $\phi_n$. The ability to select specific interaction terms makes the platform capable of realizing a large number of geometries. For more details, we refer the reader to Refs. \cite{Manovitz2020, Manovitz2022, Shapira2023}.

As discussed previously, the basic building block of the nonreciprocally coupled free system
is three site rings with dissipation. This can be generalized to spin systems \cite{Metelmann2015}.
For the spin system, 
the Hamiltonian 
takes the form
\begin{align}
\hat{H}  = e^{i\phi/3} (J\hat{\sigma}^{\dagger}_{1}\hat{\sigma}_{2} + J'\hat{\sigma}^{\dagger}_{2}\hat{\sigma}_{3}  + J' \hat{\sigma}^{\dagger}_{3}\hat{\sigma}_{1})  + {\rm h.c.}, \label{eq:experimental_ham}
 \end{align}
 which can be obtained from Eq. (\ref{eq:experimental_ham}) by setting $\Omega_1 = J'$ and $\Omega_2 = J$, corresponding to nearest-neighbor (NN) and next-nearest-neighbor (NNN) interactions, as well as $\phi_1  = \phi_2 = \phi/3$. This set-up was realized experimentally in Ref. \cite{Shapira2023} for the case of $J = J'$, with Refs. \cite{Manovitz2020,Shapira2023} also highlighting that a ladder of such rings can be realized by applying this procedure to a longer chain of $L$ sites (with $\Omega_{n>2} = 0$), as shown in Fig. \ref{fig:experimentaldiagram}(c). 

Here we propose to add dissipation to this ladder in order to implement a chain of nonreciprocally interacting spins, inspired by the methods described earlier. 
In particular, we 
propose to implement
loss $\hat{L} = \sqrt{\zeta} \hat{S}^-$ applied to a single site in each ring, while tuning the phase $\phi$ to facilitate the necessary interference effects.
Our proposed setup is schematically illustrated in Fig. \ref{fig:experimentaldiagram}(d).
From perturbation theory analogous to what has been done for free fermions above, 
the jump operator is given by  $\hat{L}_j = \sqrt{\kappa} (\hat{\sigma}_{j} + e^{i\phi}\hat{\sigma}_{j+1})$, 
with $\kappa={\mathcal O}(J'^2/\zeta)$, which reproduces the model we analyzed in the main text.
We note that the ladder of Aharanov-Bohm rings in Fig. \ref{fig:experimentaldiagram}(c) features undesirable couplings between the lower rung of auxiliary sites. However, according to perturbation theory, we estimate 
a shift $\delta \kappa = \mathcal{O}(J'^2 J / \zeta^2)$. This is small since we have assumed $J'/\zeta \ll 1.$ 
This implies $\delta \kappa  \ll \kappa$ 
which is sub-dominant and can therefore be neglected. Higher-order contributions are even further suppressed.
We note that, since the exponents we observe do not depend on the anisotropy $\Delta,$ $\hat{\sigma}^z_j \hat{\sigma}^z_{j+1}$ interactions in the Hamiltonian need not be implemented to obtain our core results.

 We also propose to add spin pumping $\hat{L} = \sqrt{\Gamma} \hat{S}^+$. For the implementation of both spin pumping and loss, we consider the energy level structure of Strontium ions, shown in Fig. \ref{fig:experimentaldiagram}(b) \cite{Manovitz2022}, along with the associated wavelengths for the various possible transitions. 
We first note that the qubit degrees of freedom are realized in long-lived $5S_{1/2}$ and $4D_{5/2}$ orbital states. The $5P$ orbital states are comparatively short-lived and can be used to mediate optical pumping and absorption into the qubit states \cite{Manovitz2022}. Since only the sites in the nonreciprocal chain require pumping, while sites in the auxiliary chain require loss (as described above), these processes never have to be implemented simultaneously on any individual ion.

\subsection{Free fermion results}
In this section, we derive the scaling exponents $\{z,\nu,\alpha\} = \{2, 1/2, 1/2\}$ for the free fermion model defined by Eqs. (\ref{eq:HatanoHop}) of the main text. Since the density matrix is Gaussian, the system is fully characterized by the two-point correlation matrix $\langle \hat c_m^\dagger \hat c_n \rangle$, with dynamics given by \cite{Mcdonald2022}
\begin{align}
 i \partial_t \langle c_m^{\dagger}& c_n \rangle  =   \nonumber \\ &  \frac{1}{2}(J- i \kappa e^{-i\phi}) \langle c_m^{\dagger} c_{n-1} \rangle + \frac{1}{2}(J - i \kappa e^{i\phi})\langle c_m^{\dagger} c_{n+1} \rangle \nonumber  \\&    
 - \frac{1}{2}(J+ i\kappa e^{-i\phi})\langle c_{m+1}^{\dagger} c_{n} \rangle  
 -\frac{1}{2}(J+i\kappa e^{i\phi})  \langle c_{m-1}^{\dagger} c_{n} \rangle   \nonumber \\&          - i2(\kappa +\Gamma) \langle c_m^{\dagger} c_n \rangle +  i2\Gamma  \delta_{mn}
, \label{SMeq: correlation matrix}
\end{align} 
which can be readily obtained by deriving the Heisenberg equation of motion from the Lindblad master equation of this system.

Below, we focus on the regime $L\gg J/\Gamma$ and work in the thermodynamic limit $L\rightarrow \infty$. As argued in the main text, in this regime, boundary conditions are not going to play a crucial role in bulk physics, even in the presence of the Liouvillian skin effect (LSE). 
Indeed, this is demonstrated in the main text and in Ref.~\cite{Mcdonald2022},
where the free fermion system exhibits \textit{identical} steady-state profiles both in periodic (PBC) and open boundary conditions (OBC) except in the vicinity of the edge. 
This is because the system does not care about the boundary condition unless the wave packet hits the edge, 
which occurs only if the wave propagates (or diffuses) to the edge before it dampens out. As long as the damping rate of the wave packet is finite (which would be the case for finite $\Gamma>0$), a generic wave packet generated by the pumping would not reach the edge in the thermodynamic limit $L\rightarrow\infty$, making the boundary conditions irrelevant in our setup, irrespective of the presence of LSE.

Taking advantage of this, we consider PBC, for which translation invariance makes the momentum $k$ a good quantum number.
Fourier transforming Eq.~\eqref{SMeq: correlation matrix}, one finds that the momentum distribution function $n_k = \frac{1}{L}\sum_{mn} e^{i k (m-n)} \langle \hat c^\dagger_m \hat c_n\rangle$ is given by
\begin{align}
&\partial_t  n_{k} =  - 2\big\{\kappa(1+ \cos(\phi+ k))  +  \Gamma   \big\} n_{k} +  2\Gamma .  
\label{eq:nequat}
\end{align}
This has the general solution 
\begin{align}
 n_k (t)= & c(k)  e^{- 2\big(\kappa(1+ \cos(\phi+ k)+\Gamma \big)t} \nonumber \\ &+ \frac{\Gamma}{\kappa(1+ \cos(\phi+ k))+\Gamma},
\end{align}
where $c(k)$ is a constant of integration that depends on the initial conditions. The mean particle density  is then obtained by integrating over momentum space:
\begin{align}
\bar{n} (t)   =& \int_{-\pi}^{\pi} \frac{dk}{2\pi} c (k)  e^{- 2\big(\kappa(1+ \cos(\phi+ k))+\Gamma \big)t} \nonumber  \\& + \int_{-\pi}^{\pi} \frac{dk}{2\pi}\frac{\Gamma}{\kappa(1+ \cos(\phi+ k))+\Gamma} \label{eq:contint}
\end{align}
Expanding around the least damped mode $k=k_* = \pi - \phi$ and extending the integral bounds to $\pm \infty$, one obtains,
\begin{align}
 \bar{n} (t)   =& \int_{-\infty}^{\infty} \frac{d\Delta k}{2\pi} c \,  e^{-\big(\kappa(\Delta k)^2+2\Gamma \big)t} \nonumber  \\& +  \int_{-\infty}^{\infty} \frac{d\Delta k}{2\pi} \frac{2\Gamma/\kappa}{(\Delta k  + i \sqrt{2\Gamma/\kappa})(\Delta k  - i \sqrt{2\Gamma/\kappa})} \nonumber \\
 = & ~\frac{ c \,  e^{-2\Gamma t} }{2\sqrt{\pi\kappa t}} + \sqrt{\frac{\Gamma}{2\kappa}} ,
\label{eq:density_hatano}
\end{align}
where we have also assumed that $c(k)\sim c$ is independent of $k$ in the vicinity of the 
least damped 
mode. 
The steady-state mean density $\bar n(t\rightarrow\infty)$ is given by
$\bar{n}(t\rightarrow\infty) \sim \Gamma^\alpha =\Gamma^{1/2}$, giving the scaling characterized by the exponent $\alpha = 1/2$ demonstrated in Fig. \ref{fig:steadystate}(a) of the main text in the 
asymptotic regime.
One also sees from the first term of Eq.~\eqref{eq:density_hatano} that $\Gamma$ sets a relaxation time-scale $t_r \sim 1/\Gamma$. 
In the limit $\Gamma\rightarrow 0$,  this timescale diverges and
the relaxation reduces to a power law $\bar{n} (t)  \sim t^{-\chi}= t^{-1/2}$ characterized by the exponent $\chi = 1/2$, as demonstrated
in Fig. \ref{fig:chiral}(b) of the main text.

We now derive the steady-state correlation functions. 
The steady-state correlation function $(\partial_t\langle \hat c^\dagger_m \hat c_n\rangle=0)$ is given from Eq.~\eqref{SMeq: correlation matrix} as
\begin{align}
\langle \hat{c}_m^{\dagger} \hat{c}_n \rangle
= 
\Gamma \int^{\infty}_{-\infty} \frac{d\omega}{2\pi}  \bra{n} \frac{1}{\omega \bm{1} -\bm{H}_{\text{eff}}} \frac{1}{\omega \bm{1} -\bm{H}^{\dagger}_{\text{eff}}} \ket{m},
\end{align}
where 
\begin{eqnarray}
    &&(\bm H_{\rm{eff}})_{nm}=\frac{J}{2}  (\delta_{n m+1} + \delta_{n+1 m} )
    -\frac{i}{2}\Gamma\delta_{nm}
    \nonumber\\
    &&-\frac{i}{2}\kappa
    (\delta_{n m} 
    + \delta_{n+1,m+1}
    +e^{i\phi}\delta_{n+1,m}
    +e^{-i\phi} \delta_{n,m+1} )
    \nonumber\\
\end{eqnarray}
is the effective Hamiltonian (that is in fact different from the conditional Hamiltonian, see Ref. \cite{Mcdonald2022}) in the first quantized form, and $\ket{m} = \hat c_m^\dagger\ket{0}$ is a position eigenstate (where $\ket{0}$ is a vacuum state).
After performing the Fourier transformation, one arrives at
\begin{eqnarray}
&&\langle c_m^{\dagger} c_n \rangle =  \frac{1}{L}\sum^{B.Z.}_{k}  \Gamma  e^{ik(n-m)} 
\nonumber\\ && \  \times  \int^{\infty}_{-\infty} \frac{d\omega}{2\pi} \bra{k} \frac{1}{\omega \bm{1} -\bm{H}_{\text{eff}}}   \ket{k}  \bra{k}\frac{1}{\omega \bm{1} -\bm{H}^{\dagger}_{\text{eff}}}\ket{k},
\end{eqnarray}
where $\ket{k}=\hat c_k^\dagger \ket{0}$ is a momentum eigenstate. Using the fact that the effective Hamiltonian is diagonal in momentum space, i.e.
\begin{align}
\bm{H}_{\rm eff} = \big(J \cos k - i\kappa(1 + \cos(\phi + k) )- i \Gamma \big) \bm{1},
\end{align}
the correlation function reduces to
\begin{align}
\langle \hat{c}_m^{\dagger} \hat{c}_n \rangle = & \frac{1}{L} 
\sum^{B.Z.}_{k}\Gamma  e^{ik(n-m)}  \int^{\infty}_{-\infty} \frac{d\omega}{2\pi} 
\nonumber\\ 
& \times \frac{1}{\omega - J \cos k + i\kappa(1+\cos(\phi + k)) + i\Gamma} \nonumber \\ & \times \frac{1}{\omega - J \cos k - i\kappa(1+\cos(\phi + k)) - i\Gamma} \nonumber\\
 = & \frac{1}{L}\sum^{B.Z}_{k} \frac{1}{2}\Gamma  e^{ik(n-m)}  \frac{1}{\kappa(1+\cos(\phi + k)) + \Gamma}.
\end{align}
As in the above, by expanding around the least damped mode $k=k_*=\pi - \phi$, we obtain,
\begin{align}
&\langle \hat{c}_m^{\dagger} \hat{c}_n \rangle 
=  \int_{-\infty}^{\infty} \frac{d\Delta k}{2\pi}  \frac{\Gamma}{\kappa}  e^{i(\Delta k-\phi + \pi)
(n-m) } 
\nonumber  \\ & \times \frac{1}{(\Delta k - i \sqrt{2\Gamma/\kappa})(\Delta k + i\sqrt{2\Gamma/\kappa}) },
\end{align}
where once again we have extended the integral bounds to $\pm \infty$.
Performing the contour integral leads to
\begin{align}
\langle \hat{c}_m^{\dagger} \hat{c}_n \rangle 
= \frac{1}{2\sqrt{2}}\sqrt{\frac{\Gamma}{\kappa}}  e^{-\sqrt{2\Gamma/\kappa}|n-m|
+ i (\pi -\phi) 
(n-m)
},
\end{align}
showing that the correlation length is given by $\xi=\sqrt{\kappa/\Gamma}$. This implies $\xi \sim \Gamma^{-\nu}$ with $\nu = 1/2$. A comparison of (\ref{eq:density_hatano}) with the scaling ansatz Eq. (\ref{eq:densityscaling}) in the main text indicates that $\chi = \alpha / (\nu z), $ which implies that $z = 2.$ This concludes our derivation of the critical exponents $\{z,\nu,\alpha\}  =  \{2,1/2,1/2 \}$ for the free fermion system. 

Finally, we consider the impact of finite size effects on the scaling of the steady state density in the red region of Fig. \ref{fig:steadystate}(a),  under periodic boundary conditions. 
In this instance, the integral (\ref{eq:contint}) should be replaced by a sum over discrete momenta,
\begin{align}
\bar{n}(t\rightarrow \infty) & = \sum_{k}^{B.Z.} \frac{\Gamma/\kappa}{1+\cos(\phi+ k) + \Gamma/\kappa}.
\end{align}
In the limit $\Gamma \rightarrow 0,$ $\Gamma/\kappa$ becomes very small compared to the gap, which will be $\mathcal{O}((2\pi/L)^2)$. We can therefore ignore $\Gamma/\kappa$ in the denominator, leaving
\begin{align}
\bar{n}(t\rightarrow \infty)  & \approx \frac{\Gamma}{\kappa} \sum_{k}^{B.Z.} \frac{1}{1+\cos(\phi+ k)} 
\end{align}
Hence, 
$\bar{n} \sim \Gamma,$ ($\alpha' = 1)$, provided $\Gamma/\kappa \leq \mathcal{O}(1/L^2)$. 
We remark that this is different from the OBC case, where this scaling emerges at a threshold that is linear in the system size, i.e. $\Gamma/\kappa = \mathcal{O}(1/L)$.

\subsection{Spin wave dark state for XXZ}
In this section, we show that in the thermodynamic limit (where momentum $k$ is a continuous variable), our model has
a spin wave mode that is a dark state. 
As discussed in the main text, the dissipator given by Eq. (\ref{eq:dissipator}) in the main text vanishes when $k=k_*$.
This implies that a state $\ket{D_k}\equiv \hat{S}_k^+ \ket{\Downarrow}$ does not experience any dissipation when $k=k_* = \pi-\phi$, where the operator $\hat{S}_k^+ =\frac{1}{\sqrt{L}}\sum_{j=1}^L e^{ikj} \hat{S}_j^+$ creates a spin-wave mode with momentum $k$. In addition, $\ket{D_{k^*}}$ is also an eigenstate of the Hamiltonian $\hat H$, because 
no scattering would occur in the one-magnon sector (where the state $\ket{D_{k^*}}$ lives), and therefore, the momentum would stay preserved. Indeed, by operating on this state with
the XXZ Hamiltonian (Eq. (\ref{eq:XXZ}) of the main text), one finds,
\begin{align} 
\hat{H} \ket{j} = \frac{J}{2}\ket{j-1} + \frac{J}{2}\ket{j+1} + \frac{(L-4)J\Delta}{4}\ket{j},
\end{align}
which has $\ket{D_k} = \frac{1}{\sqrt{L}}\sum_j e^{ijk}\ket{j}$ as an eigenstate 
with eigenvalues $\frac{L-4}{4}J\Delta + J\cos k$, including the least damped mode $k=k_*$.
Combining these two properties yields $\mathcal{L}[\ket{D_{k^*}}\bra{D_{k^*}}]=0$, meaning that $\ket{D_{k^*}}$ is a dark state.

\subsection{$U(1)$ symmetry and single-magnon eigenstate}

Here, we show that the U(1) symmetry is a necessary and sufficient condition for the single-magnon spin wave state $\ket{D_{k}}=\hat{S}_k^+ \ket{\Downarrow}$ satisfying $\hat S_k^z\ket{D_{k}} = S_k^z\ket{D_{k}}$ (where $S_k^z$ is an eigenvalue of $\hat S_k^z$) 
to be a simultaneous eigenstate of a translationally invariant Hamiltonian $\hat H=\sum_k\hat h_k$ with periodic boundary conditions.
A single-magnon momentum state $\ket{D_{k}}$ is simultaneously an eigenstate of the Hamiltonian if $[\hat h_k,\hat S_k^z]=0$, implying that $U(1)$ symmetry of the Hamiltonian is a  necessary condition.
The sufficient condition can also be proven as follows.
In the fermionic picture, the Hamiltonian can be expressed as $\hat{H} = \sum_k \varepsilon_k \hat c_k^\dagger \hat c_k 
+\sum_{k,k',q} u_{k,k',q} \hat c_{k+q/2}^\dagger\hat c_{-k+q/2}^\dagger \hat c_{k'+q/2}\hat c_{-k'+q/2} + \cdots$, where ``$\cdots$'' represents higher-order interactions terms. 
$U(1)$ symmetry of the Hamiltonian $\hat H$ implies that the eigenstates of $\hat H$ are Fock states.
In the single-occupied Fock state sector (corresponding to the single-magnon sector), the interaction between the fermions is absent and the Hamiltonian can be written as 
$\hat{H} = \sum_k \varepsilon_k \hat c_k^\dagger \hat c_k=\sum_k\varepsilon_k\hat S_k^+ \hat S_k^-$.
This Hamiltonian satisfies $\hat{H}\ket{D_{k}}=\varepsilon_k\ket{D_{k}}$ and therefore, the sufficient condition is also proven.

\subsection{Criticality and decay channels to dark states}
We now consider the role of dark states in determining the critical properties of the spin model. Consider an arbitrary two magnon state $\ket{\psi_2}$, expressed in the general form
\begin{align}
\ket{\psi_2} = \sum_{mn} a_{mn} \ket{m,n} = \sum_{mn} a_{mn} \hat{S}_m^{+}\hat{S}_n^{+}\ket{\Downarrow} 
\label{eq:arbsup}
\end{align}
where the indices $m,n$ label the position of the spin excitations and $a_{mn}$ are arbitrary coefficients, with $a_{mn} = a_{nm}$. The action of the jump operator $\hat{L}^l_j$, given in Eq. (\ref{eq:dissipator}), on the state (\ref{eq:arbsup}) yields a one magnon state $\ket{\psi_1}$, given by 
\begin{align}
\ket{\psi_1} & = \hat{L}^l_{j} \ket{\psi_2} \nonumber \\ &  = \sqrt{\kappa}( \hat{S}^-_j +  e^{i\phi} \hat{S}^-_{j+1} )\sum_{mn} a_{mn} \ket{m,n}
\nonumber  \\ &  = \sqrt{\kappa}\sum_{mn} a_{mn} \big( \hat{S}^-_j \ket{m,n}+  e^{i\phi} \hat{S}^-_{j+1} \ket{m,n}\big) 
\nonumber  \\ &  = \sqrt{\kappa}\sum_{mn} a_{mn} \Big( \delta_{j,m}\ket{n}  + \delta_{j,n}\ket{m} \nonumber \\ & ~~~~~~~~~ +  e^{i\phi} (\delta_{j+1,m}\ket{n}  + \delta_{j+1,n}\ket{m} ) \Big) \nonumber  \\&
   = 2\sqrt{\kappa} \sum_n \big(a_{jn}  +  e^{i\phi} a_{j+1n}   \big) \ket{n}, \label{eq:finaltwomag}
\end{align}
where $\ket{n}$ denotes a state with  a single magnon at position $n$. To reach the last line we performed a  summation and used $a_{mn} = a_{nm}$. Setting (\ref{eq:finaltwomag}) equal to the spin wave state $\ket{k} = \frac{1}{\sqrt{L}}\sum_{n=1}^L e^{ikn} \ket{n}$, we obtain
\begin{align}
a_{jn} +  e^{i\phi} a_{j+1n} =\frac{1}{\mathcal{N}} e^{ikn} , \label{eq:magnon_contraint}
\end{align}
where $\mathcal{N}$ is a normalization constant. This includes the specific case of the dark state, for which $k = k_* = \pi - \phi.$ This places a unique constraint on the coefficients $a_{mn}$ that are required in order to arrive at a particular spin wave state after a quantum jump. In particular, once $a_{j1}$ is fixed all terms $a_{jn}$ are specified by applying (\ref{eq:magnon_contraint}) recursively. The apparent ambiguity in `choosing' $a_{j1}$ is removed by the normalization condition (up to a phase).

We now demonstrate that this decay of a two-magnon state via quantum jumps is the only way to reach the state $\ket{D_{k_*}}$. 
Due to the U(1) symmetry of the XXZ Hamiltonian, as shown in the section ``$U(1)$ symmetry and dark state'', the momentum modes form an eigenbasis of the single magnon sector. This ensures that there are no transitions between momentum modes.  
Putting these facts together, Eq. (\ref{eq:magnon_contraint}) provides the requirement for reaching $\ket{D_{k_*}}$ during the time evolution. The freedom to choose the index $j$ indicates that there are $L$ possible transitions to this state from linearly independent states in the two-magnon sector, corresponding to the possible locations where the quantum jump may occur.

This implies that a basis for two magnon states can be constructed in which the remaining $\mathcal{O}(L^2)$ basis states decay to the vacuum over time via two jumps. 
We therefore conclude that there are macroscopically more ways to arrive at the vacuum dark state $\ket{\Downarrow}$ than the spin wave dark state $\ket{D_{k_*}}$. This implies that for large systems the vacuum will be the steady state. The exception is the scenario in which the initial state has a large overlap with $\ket{D_{k_*}}$, since this is preserved throughout the time-evolution. However, for generic initial states, this overlap is typically exponentially small. Modes in the vicinity of the dark state, i.e. in some momentum shell $\Delta k$ from the dark state such that $k = k_* + \Delta k$, therefore play the role of determining the long time relaxation towards the vacuum dark state.  The slow relaxation of these modes is directly implied from the dissipator in Eq. (\ref{eq:momemtlind}) of the main text.

For free fermions, this picture can be stated analytically.  Like the spin system, the vacuum $\ket{0}$ is a trivial dark state with a macroscopically large number of decay channels, and there is the momentum dark state at $k = k_* = \pi - \phi$, where $\ket{k} = \hat{c}^{\dagger}_k\ket{0},$ and  $\hat{c}_k^+ =\frac{1}{\sqrt{L}}\sum_{j=1}^L e^{ikj} \hat{c}_j^{\dagger}$. The momentum space retarded Greens function \cite{Mcdonald2022} is given by 
\begin{align}
G^R(\omega,k) = \frac{1}{\omega - J \cos k + i\kappa(1 + \cos(\phi + k) ) +  i \Gamma }, \label{ref:Greensfunc}
\end{align}
where it can be seen that when $\Gamma = 0$ the dissipative gap $-\kappa(1 + \cos(\phi + k) )$ vanishes at $k = k_* = \pi - \phi.$ 
This Green's function encodes the response following a perturbation to the steady state configuration, which through the macroscopic number of decay channels is guaranteed to be the vacuum state $\ket{0}$. Clearly, the asymptotic features will be determined by modes in the vicinity of $k_*$, as can be seen by expanding around this mode using $\cos(\phi + k) \approx -1 + \frac{1}{2}(\Delta k)^2$. For the spin system, we anticipate a similar form to  (\ref{ref:Greensfunc}), albeit with a self-energy contribution $\Sigma(k).$ A full treatment within the framework of Keldysh field theory is highly non-trivial and is left for future work.

\subsection{Liouvillian spectrum}
Here, we demonstrate the presence of the LSE for the dissipative XXZ system by computing the eigenvectors and eigenvalues.
We first discuss the case in which $\Gamma = 0.$ For loss-only models the Liouvillian can be expressed in a block-triangular form \cite{Albertini1996,Torres2014,Yoshida2020,Buca2020,Nakagawa2021,Yoshida2023}, which allows the Liouvillian spectrum to be determined entirely by the conditional Hamiltonian $\hat H_{\rm cd}$.
The eigenvalues of the Liouvillian are given by $\lambda_{(ij)} = \varepsilon_i + \varepsilon_j^*$, where $\varepsilon_i$ is the spectrum of $-i\hat{H}_{\rm cd}$. 
When $\Delta = 0$ the conditional Hamiltonian is the same as the free model, see Eqs. (\ref{eq:fermcond}) and (\ref{eq:hcondspinJW}) of the main text. 
The nonreciprocal case, therefore, matches the spectrum of the Hatano--Nelson model up to a shift, 
which has analytically been shown to have
spectral sensitivity to boundary conditions \cite{Lee2016,Yao2018}; see the insets in Fig. \ref{fig:asymprop}(b) of the main text. 
However, 
at finite $\Gamma>0$,
as the Liouvillian cannot be expressed in a  block-triangular structure, the spectrum of the spin system cannot be calculated from the free system.
We, therefore,
resort to numerical methods.

\begin{figure}
\begin{tabular}{c|c|c|c|c} 
\hline 
 $\Delta$   \,  & ~~~~~~ 0 ~~~~~~ & ~~~~~~~ 1 ~~~~~~~ & ~~~~~~ 1 ~~~~~~~ & ~~~~~~ 1 ~~~~~~~ \\ \hline
 $\Gamma/\kappa$ \, & 0 & 0 & 0.056 & 0.11 \\ \hline 
\end{tabular}

\vspace{-0.3cm}

    \centering
     \subfloat[\hspace{1.60cm}(a)\hspace{1.40cm} (b) \hspace{1.40cm} (c) \hspace{1.35cm} (d)]{\includegraphics[width = \columnwidth]{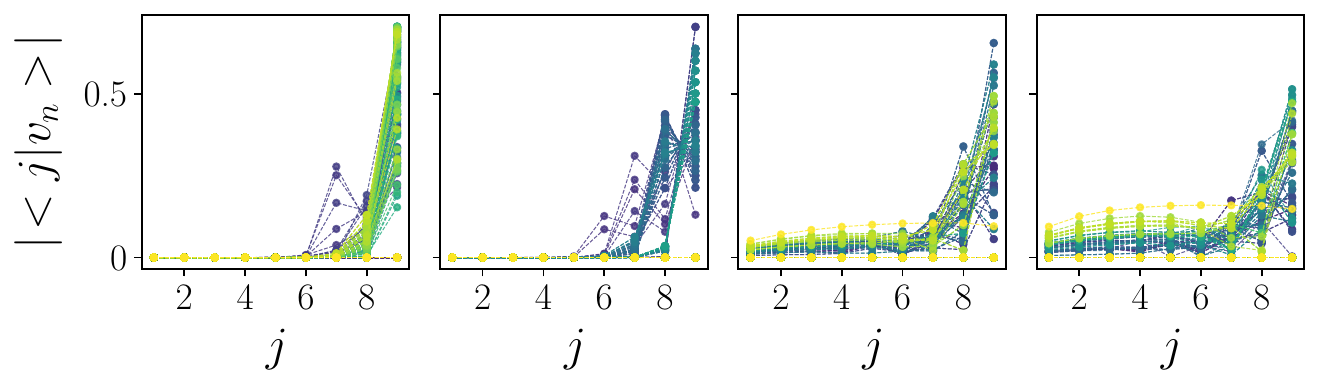}} 

\vspace{-0.3cm}
  \centering
  \subfloat[\hspace{1.60cm}(e)\hspace{1.5cm} (f) \hspace{1.39cm} (g) \hspace{1.40cm} (h)]{\includegraphics[width = \columnwidth]{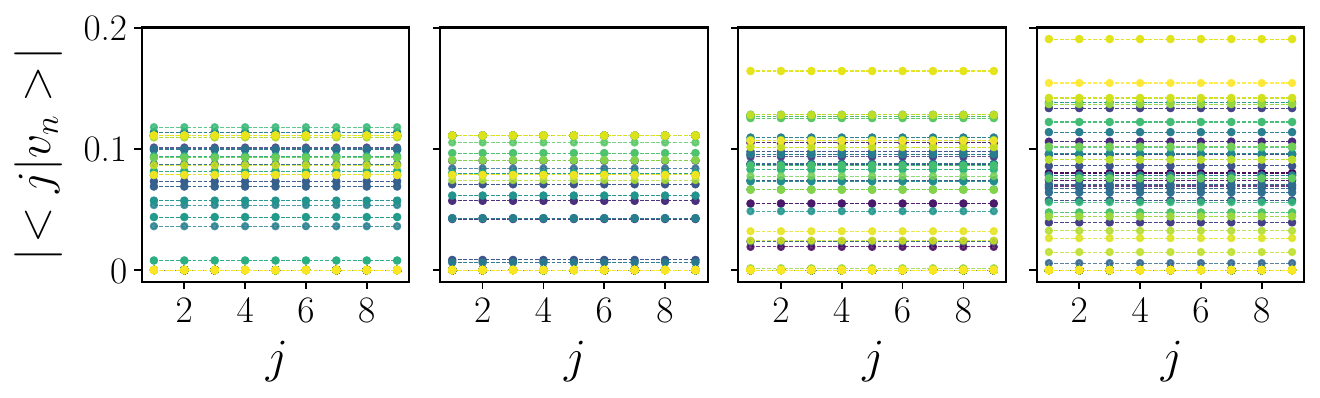}}

\vspace{-0.3cm}
  \centering
         \subfloat[\hspace{1.60cm}(i)\hspace{1.55cm} (j) \hspace{1.45cm} (k) \hspace{1.37cm} (l)]{\includegraphics[width = \columnwidth]{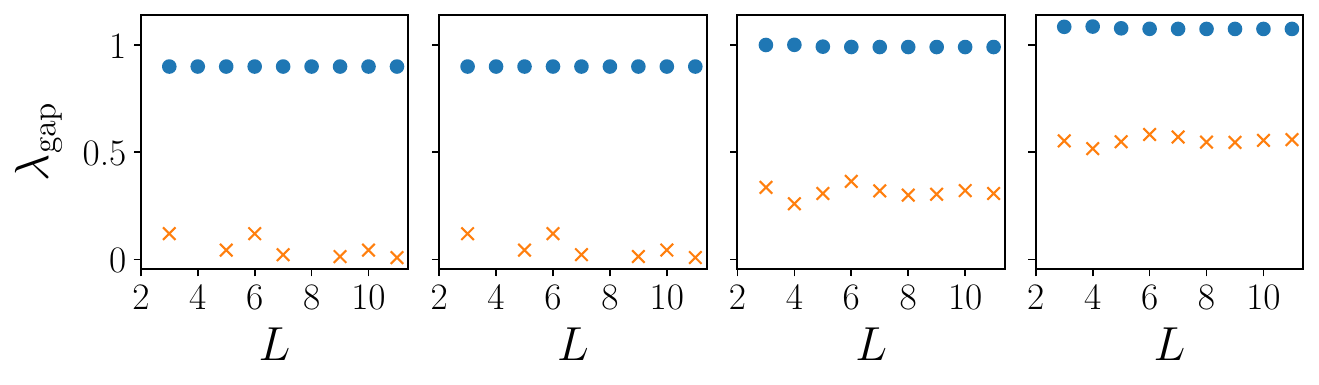}}

       \vspace{-0.3cm}
   
        \subfloat[\hspace{1.60cm}(m)\hspace{1.40cm} (n) ]{\includegraphics[width = 0.54\columnwidth,left]{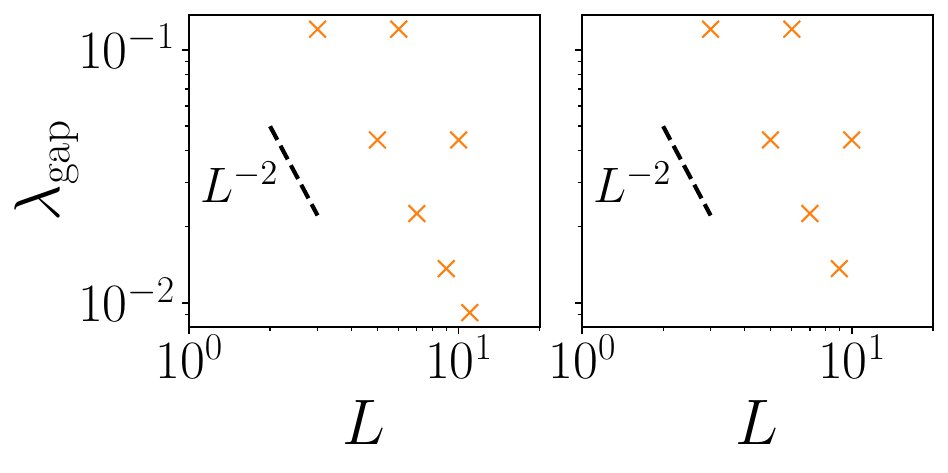}}
    \caption{
    Eigenvectors and eigenvalues of an $L = 9$ nonreciprocal ($\phi = -\pi/2$) XXZ spin chain for a variety of parameters (see column labels at the top) and with $J/\kappa = 1.11$. 
   (a)-(d) Spatial support of eigenvectors $|\braket{j}{v_n}|$ for OBC. 
   The data represents the 200 eigenstates with the largest $\text{Re} [\lambda_n]$. 
   The eigenvectors are colored according to the size of the corresponding eigenvalue; those with 
   a larger real part are bright colors while those with a smaller real part are darker. (e)-(h) $|\braket{j}{v_n}|$ for PBC. (i)-(l) Liouvillian gap $\lambda_{\rm gap}$ vs system size $L$ for OBC (dots) and PBC (crosses).
   (m)-(n) The PBC data is also plotted on a log-log plot for the cases with $\Gamma = 0$, demonstrating that the gap is approximately closing with $1/L^2$.
   } 
    \label{fig:spectrum_cases}
\end{figure}

Figure \ref{fig:spectrum_cases} shows the spectral properties for a range of system parameters, 
obtained numerically via exact diagonalization, which can be obtained using standard sparse matrix eigensolvers.
Let $\ket{v_n}$ be the vectorized representation of the right eigenvectors (which are matrices) satisfying $\mathcal{L}[\hat{\rho}^R_n] = \lambda_n \hat{\rho}^R_n$. Here, the eigenvalues are labeled in descending order of their real part
$0=\text{Re}[\lambda_0] \geq  \text{Re}[\lambda_1] \geq  ..\geq \text{Re}[\lambda_{4^L}]$. 
Fig. \ref{fig:spectrum_cases}(a)-(d) shows the spatial support $|\braket{j}{v_n}|$ under OBC for a variety of system parameters. Here, the first 200 eigenstates  $\ket{v_0},..., \ket{v_{199}}$ (i.e, the eigenstates with the first 200 slowest modes) are plotted. 
In all cases, the localization is clearly visible at the right-hand boundary. The localization length of the OBC eigenvectors is insensitive to changes in $L$ (not shown), implying that the localization of eigenmodes would persist in the thermodynamic limit.
Fig. \ref{fig:spectrum_cases}(e)-(h) shows the results under PBC. In stark contrast to the OBC results, all eigenvectors are 
spatially uniform.
Fig. \ref{fig:spectrum_cases}(i)-(l) shows the Liouvillian gap $\lambda_{\rm gap} = |\text{Re} [\lambda_1]| $ vs system size $L$ under OBC and PBC respectively.
When $\Gamma = 0$, the gap closes with $1/L^2$ (see Fig. \ref{fig:spectrum_cases}(m)-(n)) under PBC, while for OBC  the gap does not close. The situation is qualitatively similar at finite $\Gamma,$ albeit shifted by the gap. Taken together, these plots clearly illustrate the spectral sensitivity to the choice of boundary conditions, even at finite $\Gamma$. 

\captionsetup[subfloat]{labelformat=empty}

 \subsection{Numerics}
We now provide details of
the tensor network methods employed throughout this work. We use the time-evolving block decimation (TEBD) algorithm \cite{Vidal2004} to perform numerical simulations. 
We start by moving to a vectorized representation of the Lindblad equation, with a density matrix 
$  \text{vec} [\hat{\rho}] = \ket{\rho}\label{eq:vectoriser}$. 
Operators acting on the bra and ket respectively now act on different copies of the Hilbert space $ \mathcal{H}   \otimes  \mathcal{H}$, with $\text{vec} [ \hat{A} \rho \hat{B}] =  \hat{A} \otimes \hat{B}^T  \ket{\rho}$. In this representation the Lindblad equation (\ref{eq:lindblad})  can be rewritten as $\frac{d \ket{\rho}}{dt} =  \hat{\mathcal{L}}  \ket{\rho} ,$
where 
\begin{align} \hat{\mathcal{L}}  = &  -i \Big( \hat{H}\otimes  \hat{1} - \hat{1} \otimes \hat{H}^T \Big) + \sum_{\mu} \hat{L}_{\mu} \otimes  \hat{L}_{\mu}^{*} \nonumber \\ &- \frac{1}{2}\hat{L}_{\mu}^{\dagger}\hat{L}_{\mu} \otimes \hat{1} - \frac{1}{2} \hat{1} \otimes \hat{L}_{\mu}^{T} \hat{L}_{\mu}^*  .  \label{eq:vectorlind}\end{align}
The state at time $t$ is given by $\ket{ \rho(t) }=e^{\mathcal{\hat{L}} t} \ket {\hat{\rho}(0) }\label{eq:vectorsol} $. Splitting $\hat{\mathcal{L}}$ into operators that act on even and odd bonds, $\hat{\mathcal{L}}= \hat{\mathcal{L}}_{\text{odd}} + \hat{\mathcal{L}}_{\text{even}}$, the second order trotter decomposition can be used to obtain a time-evolution operator for a single time-step $\delta$:
\begin{align}
    e^{\mathcal{\hat{L}} \delta} = e^{\frac{\delta}{2}\mathcal{\hat{L}}_{\text{even}}} e^{\delta \mathcal{\hat{L}}_{\text{odd}}} e^{\frac{\delta}{2} \mathcal{\hat{L}}_{\text{even}}}  + \mathcal{O}(\delta^3) \label{eq:gates},
\end{align}
where the error per step is $\mathcal{O}(\delta^3)$. Each exponential  in (\ref{eq:gates}) is sequentially applied to a matrix product state (MPS) representation of $\ket{\rho}$ with the doubled local Hilbert space dimension on each site. 
After each of these operations, the bond dimension of the MPS may grow and should be truncated, see Refs. \cite{Schollwock2011,Paeckel2019} for a general discussion of this procedure. 
The feasibility of these simulations is massively assisted by the sinks of spin, which ensures that the state is relatively quickly brought into the proximity of the product state $\ket{\Downarrow}$. We find that the results converge with very small errors provided the bond dimension is at least 15, as discussed below.
Since $\hat{\mathcal{L}}$ is non-Hermitian, the orthogonality of the Schmidt eigenstates  is not automatically preserved \cite{Daley2005}. For the case of action on even (odd) bonds, with $e^{\frac{\delta}{2}\mathcal{\hat{L}}_{\text{even}}}$ ($e^{\delta\mathcal{\hat{L}}_{\text{odd}}}$), our approach is to sweep from left to right. Rather than skipping the odd (even) bonds, we apply the identity matrix followed by a singular value decomposition, which re-orthogonalizes the Schmidt eigenstates on that bond. Once the end of the chain is reached, we then perform a singular value decomposition on every bond in a sweep from right to left, i.e. in the opposite direction, which enforces orthogonality of the Schmidt eigenstates on each respective bond. The procedure is then repeated for the next exponential, $e^{\delta\mathcal{\hat{L}}_{\text{odd}}}$ ($e^{\frac{\delta}{2}\mathcal{\hat{L}}_{\text{even}}}$)  as defined by (\ref{eq:gates}). 

 \begin{figure}[t]
      \subfloat{\includegraphics[width=\columnwidth]{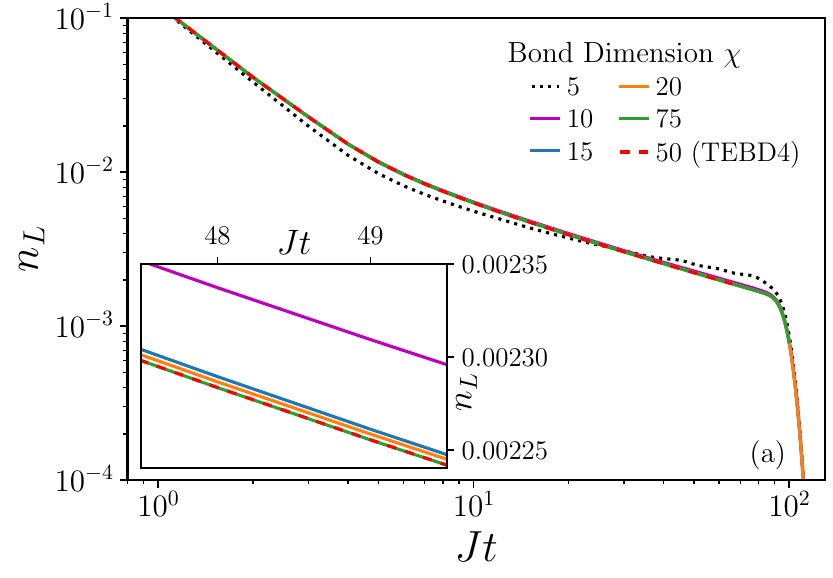}\label{fig:Error_bond_dim}}

      \subfloat{\includegraphics[width=\columnwidth]{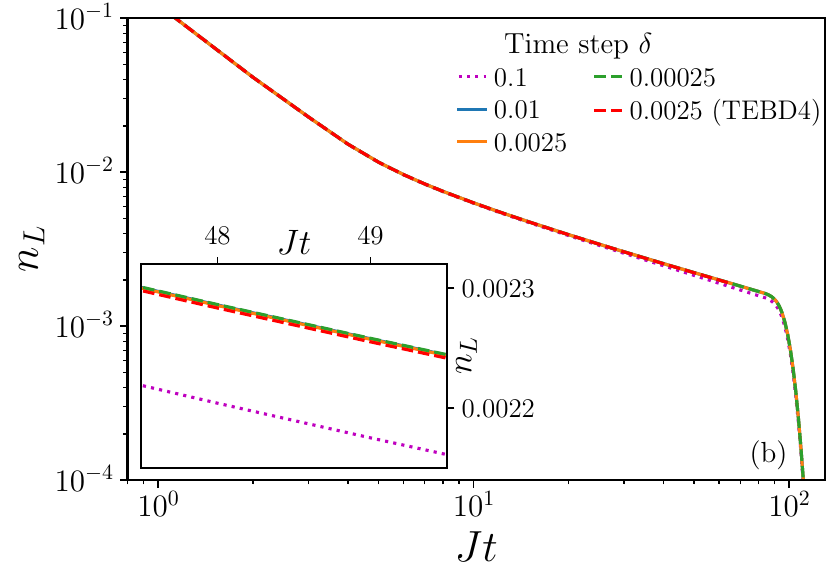}\label{fig:Error_timestep}}
      \caption{ (a) Density $n_L(t)$ vs time for the nonreciprocal ($\phi = -\pi/2$) XXZ model with $\Delta = 2$, $J/\kappa = 1$ and $L = 100$ sites. Results are shown for a number of different bond dimensions $\chi$ using the TEBD2 algorithm with fixed time-step $\delta = 0.0025$. The inset shows a zoom in of the data points. An additional comparison to TEBD4 with $\delta = 0.0025$ and $\chi = 50$ is also shown up to $Jt = 66$. This simulation, along with the $\chi =75$ TEBD2 simulation, do not run until the very end of the displayed time-interval due to limitations of the available computational resources. (b) Similar to (a) but for varying time-step $\delta$ with fixed bond-dimension $\chi = 20.$ }
     \label{fig:mps_errors}
\end{figure}
 Observables are computed using standard techniques for MPS \cite{Schollwock2011}, noting that the vectorized density matrix should be contracted with a purification of an infinite temperature state $\bra{1}$ (see Ref. \cite{Schollwock2011} for an example of such a purification), i.e. $\text{tr} (\hat{O} \hat{\rho(t)}) = \bra{1} \hat{O}_D \ket{\rho(t)}, $ where $ \hat{O}_D = \hat{O} \otimes \hat{1}.$ We have verified the algorithm against direct integration of the Lindblad equation (\ref{eq:lindblad}) for small system sizes. 
 
 In Fig. \ref{fig:mps_errors}(a) we show the density $n_L(t)$ vs time for the $\Delta = 2$ nonreciprocal ($\phi = -\pi/2)$ XXZ chain with $L = 100$ spins initialized in the fully-polarized state $\ket{\Uparrow}$. Results for a variety of bond dimensions $\chi$ are shown, with the time-step fixed as $\delta = 0.0025$, demonstrating rapid convergence with increasing bond dimension (see inset for zoomed in data). The discrepancy from the converged result is less than $\mathcal{O}(10^{-5})$ for $\chi \geq 15;$ this is less than $1\%$ of the density ($0.3\%$ for the displayed result). We therefore adopt $\chi = 15$ for the majority of simulations in the main text. We have also included  additional results for the fourth order integrator TEBD4 with  $\chi = 50$ and $\delta = 0.0025$ (data is only displayed up to $Jt = 66$ for this simulation). This algorithm reduces the time-step error to $\mathcal{O}(10^{-9})$ for the time-scales of interest. As can be seen from the inset, this agrees with the TEBD2 simulations with $\chi = 75$, suggesting that the choice of time-step for the TEBD2 results ($\delta = 0.0025$) is also sufficient. This is shown more rigorously in Fig. \ref{fig:mps_errors}(b) which presents a similar analysis but for varying time-step $\delta$, holding the bond dimension fixed at $\chi = 20$. In this case, the results change very little for $\delta \leq 0.01$. Throughout this work we therefore use $\delta = 0.0025.$
 While the results in the main text are obtained via TEBD2, the results in Fig. \ref{fig:relaxation_alldata} are obtained via TEBD4 with $\delta = 0.0025$ and a bond dimension $\chi = 15$.

\begin{figure}[t]
    \centering
     \subfloat{\includegraphics[width=\columnwidth]{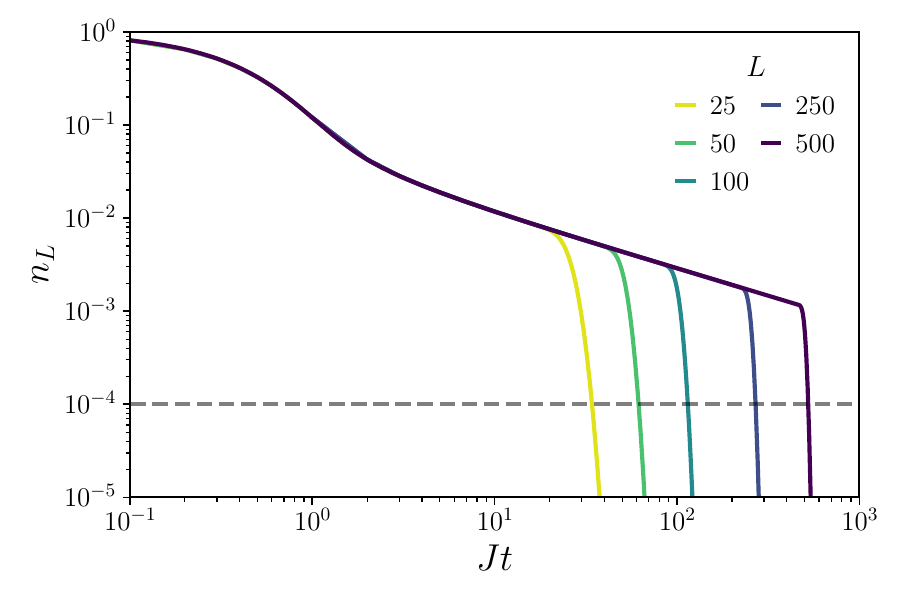}}
    \caption{Density for the site on the right edge, $n_L$, vs time for the nonreciprocal ($\phi = -\pi/2)$ XX model ($\Delta = 0$), starting from an initial state with all spins up $\ket{\Uparrow}$. Results are shown for different system sizes $L$ (legend). The dashed line indicates the cut-off density $\epsilon$ used to diagnose the relaxation time. We set $J/\kappa = 1$.}
    \label{fig:skin_effect_diagnostic}
\end{figure}

 \subsection{Diagnosing relaxation time}
This section discusses how we calculate the relaxation times shown in the inset of Fig. \ref{fig:chiral}(a) of the main text. We diagnose the relaxation time $t_r$ as the first time for which $n_j(t) < \epsilon, $ with $\epsilon$ determining the cut-off. We use $\epsilon = 10^{-4}$, although the precise choice is not important provided it is less than  $n_j(t)$ during the period of power-law decay.  In Fig. \ref{fig:skin_effect_diagnostic} we show the density $n_L(t)$ vs time for the dissipative XX model ($\Delta = 0$) with $\Gamma = 0$. The system is initially in the fully-polarized state with all spins up $\ket{\Uparrow}$, and we display results for a range of system sizes. The dashed line indicates the threshold $\epsilon = 10^{-4}$. In general we choose a site $j$ near the right edge of the system, since this maximizes the amount of time spent in power-law decay, leading to faster convergence with $L$. For finite $\Delta$ the sites very close to the right edge are atypical, so it best to consider sites slightly away from the boundary. The results in the inset of Fig. \ref{fig:chiral}(a) are therefore obtained using site $j = L-10$. An alternate approach is to precisely characterize the time-scale at which $n_j$ transitions from power-law to exponential decay, which gives similar results.

\subsection{Finite size scaling and initial state dependence of $\chi$}

 \begin{figure}[t]
      \subfloat{\includegraphics[width=\columnwidth]{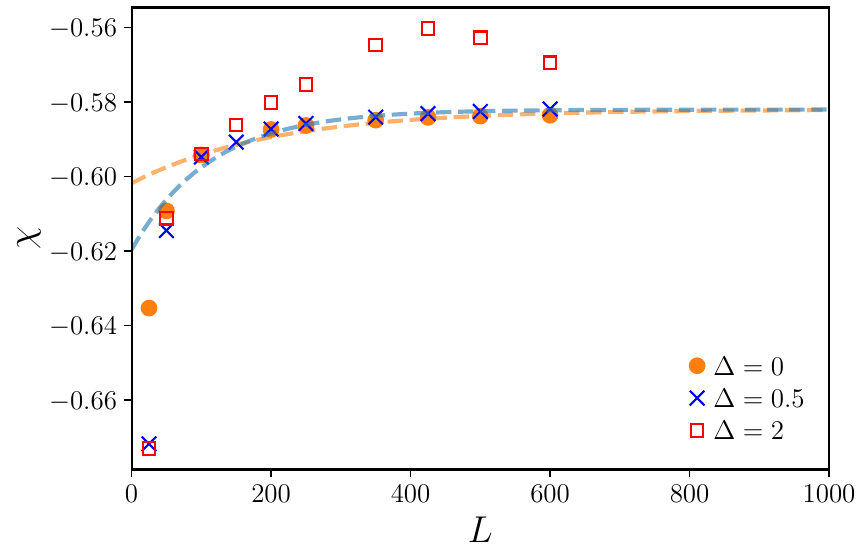}}
     \caption{
     Exponent $\chi$, defined via $n_j \sim t^{-\chi }$, vs system size for simulations initialized in the state with all spins-up $\ket{\Uparrow}$ (see Fig. \ref{fig:lse_powerlaw}). For $\Delta = 0$ and $\Delta = 0.5$ the dashed lines are fitted to system sizes $L \geq 100$ and characterize the exponential decay towards an asymptotic result. We set $J/\kappa = 1$.} 
     \label{fig:FinSize}
\end{figure}

In this section, we  perform
the finite-size scaling of the exponent $\chi$, which characterizes the power-law relaxation of the excitation density according to $n_j \sim t^{-\chi}$ at $\Gamma = 0$. We also report the initial state dependence of the exponent $\chi$.
Fig.  \ref{fig:FinSize} shows $\chi$ vs system size for the quench dynamics starting from the state with all spins up $\ket{\Uparrow}$, which is displayed in Fig. \ref{fig:chiral}(b) of the main text and Fig. \ref{fig:skin_effect_diagnostic}.
Results are displayed for different $\Delta$ values. 
For $\Delta = 0$, the exponent is extracted from the density at the boundary, i.e. $j = L$. 
For $\Delta = 0.5$ and $\Delta = 2$, $j = L - 10$ is used to avoid atypical behavior very close to the boundary. The dashed lines (fitted to the data for $L \geq 100$) give a finite-size extrapolation  to
$L \rightarrow \infty$ for the cases of $\Delta  =0$ and $\Delta = 0.5$, respectively. 
We find that the $L = 600$ data value of  $\chi = 0.582$ ($\chi = 0.583$) is close to the extrapolated values $\chi = 0.58(2)$ ($\chi  = 0.581(2)$) for $\Delta = 0$ ($\Delta = 0.5$). This suggests that the finite-size correction is quite small. 
The error is chosen to reflect the range of $\chi $ values that when fitted appear broadly compatible with the data. This process is crude and results in uncertainty that greatly exceeds the estimated numerical error for $\chi $ due to the approximations of the TEBD2 algorithm. 
For $\Delta= 2$ the finite-size correction is more significant, and we do not see convergence over this range of system sizes. However, the results do appear broadly compatible with the aforementioned cases. 

\begin{figure}[t]
    \centering
    \subfloat{\includegraphics[width=\columnwidth]{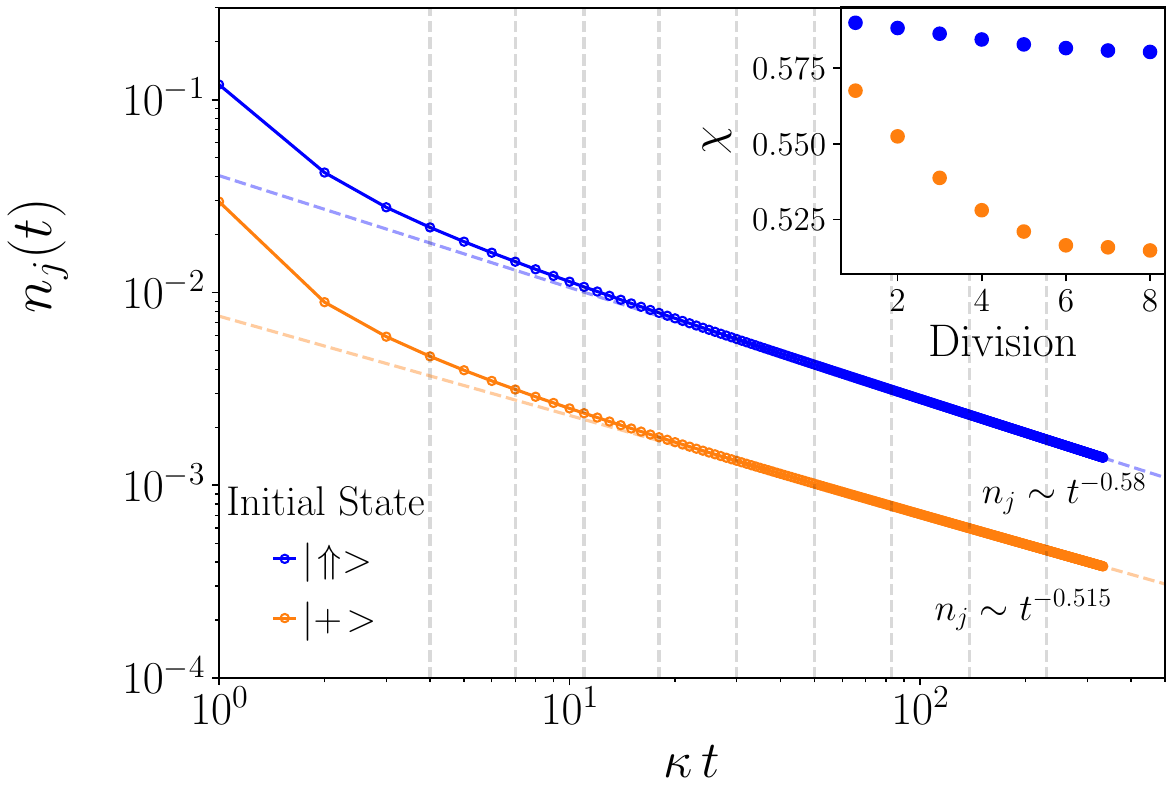}}
    
    \caption{Excitation density $n_j$ of the site in the center of the chain, $n_{L/2}(t)$, vs time for a reciprocal ($\phi = 0)$ XX model ($\Delta = 0$), starting from initial states with all spins up $\ket{\Uparrow}$ (blue) and all spins aligned in the x-direction $\prod^L_j \ket{+}_j$ (orange) respectively. We set $J/\kappa = 1$  and $L = 250.$ The main plot is split into different time divisions (vertical dotted lines). Inset: the exponent $\chi$ calculated in each division, where $n_j \sim t^{-\chi}$, demonstrating that the values for the different initial states appear not to be converging to the same value.}
    \label{fig:init_state_dep}
\end{figure}

We now demonstrate that the exponent $\chi$ is initial state dependent. Fig \ref{fig:init_state_dep} shows the decay of the excitation density $n_{L/2}(t)$ for a reciprocal ($\phi = 0$) XX model ($\Delta =0$) with $L = 250$ sites. Two different initial states are considered: the state with all spins up $\ket{\Uparrow}$ and one with all spins initially in the x-direction $\prod^L_j \ket{+}_j$. The observed $\chi$ values for the two initial states towards the end of the displayed time interval are $\chi = 0.515$ and $\chi = 0.58$. These values appear to be moderately close to convergence; the trend shown in the inset suggests that a longer simulation would only result in modest changes. 
We therefore conclude that the exponents are unlikely to converge to the same value unless there is a change in the behavior over extremely long time-scales. 
At present the reasons for the dependence on the initial conditions is not well understood. Its resolution is left for future work.

\begin{figure}[t]

     \subfloat{\includegraphics[width=\columnwidth]{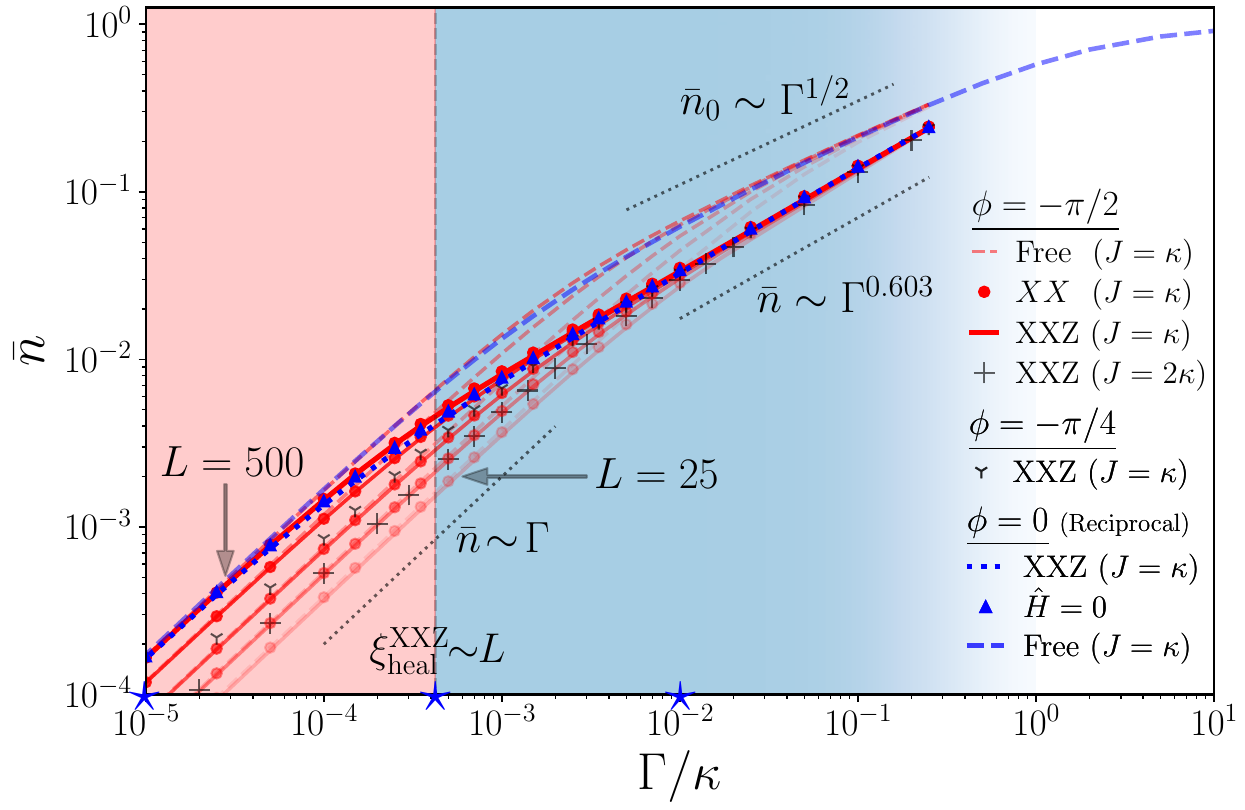} }
\caption{Excitation density $\bar{n}$ vs $\Gamma$  in the steady state for different system sizes $L$ and parameters.
Results are displayed for the following nonreciprocal ($\phi = -\pi/2$) systems:  XXZ ($\Delta= 2$), XX ($\Delta = 0$)  and free fermions for $L = \{25,50,150,250,500\}$ (light to dark), as well as XXZ with $J/\kappa = 2$ ($\Delta= 1$) for $L = 100$. For $\phi = -\pi/4$ we show XXZ with and $L = 100 $.  Reciprocal ($\phi =  0$)  XXZ ($\Delta= 2$), $\hat{H}=0$, and free fermion results are also shown, all with $L = 50$.  Various fits $\bar{n} \sim \Gamma^{\alpha}$ are displayed (discussion in the main text).}     
    \label{fig:steadystate_alldata}
\end{figure}

\begin{figure}
       \centering
   \hspace{1cm} \includegraphics[width=0.37\columnwidth]{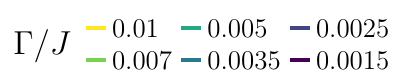}
   
       \vspace{-0.4cm}
       
     \subfloat{\includegraphics[width=\columnwidth]{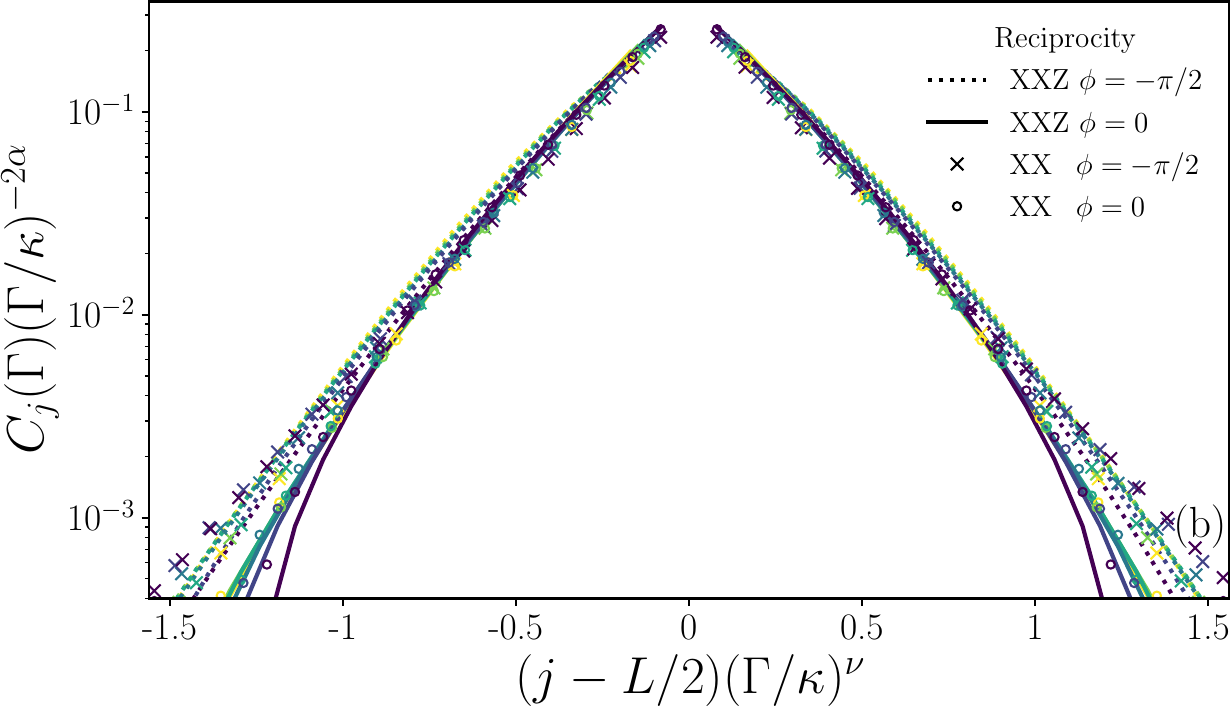} } 
    \caption{Scaled connected correlation $C_j(\Gamma)(\Gamma/\kappa)^{-2\alpha}$ vs $(j-L/2)(\Gamma/\kappa)^{\nu}$, with $\alpha=0.603$ and $\nu = 0.386$, in the steady-state for a range of $\Gamma$ values. Data is displayed for XXZ ($\Delta = 2$) and XX ($\Delta = 0$), see legend. We set $J/\kappa = 1$ and $L =250$.
    }
    \label{fig:relaxation_alldata}
\end{figure}

\subsection{Additional data}
Here, we provide additional data to further demonstrate the scaling in the main text. The additional cases were not included in the main text to preserve the figure clarity. Fig. \ref{fig:steadystate_alldata}
 shows the excitation density $\bar{n}$ vs $\Gamma$ for the XXZ system with a large number of system parameters. In comparison to  Fig. \ref{fig:steadystate}(a) of the main text, additional system size data is provided, elucidating the emergence of the scaling at larger system sizes. 

Fig. \ref{fig:relaxation_alldata} shows the scaled connected correlation $C_j(\Gamma)(\Gamma/\kappa)^{-2\alpha}$ vs $(j-L/2)(\Gamma/\kappa)^{\nu}$, see Eq. (\ref{eq:corrscale}) of the main text, for a variety of system parameters. In comparison to Fig. \ref{fig:relaxation}(b) of the main text, data is also displayed for the XX ($\Delta = 0$) cases, further demonstrating the universality.

\bibliographystyle{apsrev4-1}

\clearpage

\vfill